\documentclass[fleqn,twoside]{article}
\usepackage{espcrc2}

\newcommand{\AmS}{{\protect\the\textfont2
  A\kern-.1667em\lower.5ex\hbox{M}\kern-.125emS}}

\title{Composition of Modular Telemetry System
 with Interval Multiset Estimates}

\author{Mark Sh. Levin
\thanks{
 Mark Sh. Levin:~
 http://www.mslevin.iitp.ru;
 email: mslevin@acm.org
  }
  }

\begin{document}

\begin{abstract}
 The paper describes combinatorial synthesis approach
 with interval multset estimates of system elements for
 modeling, analysis, design, and improvement of
 a modular telemetry system.
 Morphological (modular) system design
 and improvement are considered
 as composition of the telemetry system elements (components) configuration.
 The solving process is based on Hierarchical Morphological Multicriteria Design
 (HMMD):
 (i) multicriteria selection of alternatives for system
 components,
 (ii) synthesis of  the selected alternatives
 into a resultant combination
 (while taking into
 account quality of the alternatives above
 and their compatibility).
 Interval multiset estimates
 are used for assessment of design alternatives for
 telemetry system elements.
 Two additional systems problems are examined:
 (a)
 improvement of the obtained solutions,
 (b) aggregation of the obtained solutions
 into a
 resultant system configuration.
 The improvement and aggregation processes are based on
 multiple choice problem with
 interval multiset estimates.
 Numerical examples
 for an on-board telemetry subsystem illustrate
 the design and improvement processes.

~~


\vspace{1pc}
\end{abstract}

\maketitle

\newcounter{cms}
\setlength{\unitlength}{1mm}

\section{Introduction}

 In recent decades, the significance of
 various telemetry  systems is increased
 (e.g.,
 \cite{barn11},\cite{carden02}).
 In this article, combinatorial synthesis approach
 with interval multset estimates of system elements
 is suggested for modeling, analysis, design, and improvement of
 a modular telemetry system.
 Morphological (modular) system design
 and improvement are considered
 as composition of the telemetry system element (components) configuration.
 Morphological analysis
 for system design
 is widely used many years
 (\cite{ayr69},\cite{jon81},\cite{zwi69}).
 Some recent modifications of the approach have been described in
 (\cite{lev96e1},\cite{lev06},\cite{lev09},\cite{rakov96},\cite{ritchey06}).
 Here the solving process is based on our new modification of
 Hierarchical Morphological Multicriteria Design (HMMD) approach
 (with usage of interval multiset estimates)
  (\cite{lev12morph},\cite{lev12a}):
 (i) design of a hierarchical structure
 (tree-like structure) for the designed system,
 (ii) generation of design alternatives for each
 leaf node of the system hierarchical model,
 (iii) assessment of the designed alternatives
 and their compatibility,
 (iv) multicriteria selection of alternatives for system
 components,
 (v) synthesis of  the selected alternatives
 into a resultant combination
 (while taking into account ordinal quality
 of the alternatives above and their compatibility).
 Interval multiset estimates
 are used for assessment of design alternatives for
 telemetry system elements.
 Two additional systems problems are examined:
 (a)
 improvement of the obtained solutions
 (e.g., \cite{lev06},\cite{lev10}),
 (b) aggregation of the obtained solutions
 into a resultant system configuration \cite{levagg11}.
 The improvement and aggregation processes are based on
 multiple choice problem with interval multiset estimates.

 Note, combinatorial modeling and design
 of a telemetry system has been studied in \cite{levkhod07}
 (HMMD with ordinal estimates) and
 and this work is used as a preliminary one.
 In this article, the numerical design example
 is targeted to modeling, design, and improvement
 of on-board telemetry subsystem
 while taking into account assessment
 of system components with interval multiset estimates
 \cite{lev12a}.
 The considered example corresponds to real design project.
 The example involves the following:
 tree-like structure of the subsystem,
 design alternatives (DAs) for subsystem parts/components,
 estimates of DAs and their compatibility,
 Bottom-Up design process,
 analysis and improvement of the obtained
 system solutions, and
 aggregation of the obtained solutions into the
 resultant one.
 Assessment of DAs and their compatibility is
 based on expert judgment.

\begin{center}
\begin{picture}(72,43)
\put(17,0){\makebox(0,0)[bl]{Fig. 1. Telemetric system}}

\put(00,10){\line(1,0){72}}

\put(05,10){\line(-1,-1){04}} \put(10,10){\line(-1,-1){04}}
\put(15,10){\line(-1,-1){04}} \put(20,10){\line(-1,-1){04}}
\put(25,10){\line(-1,-1){04}} \put(30,10){\line(-1,-1){04}}
\put(35,10){\line(-1,-1){04}} \put(40,10){\line(-1,-1){04}}
\put(45,10){\line(-1,-1){04}} \put(50,10){\line(-1,-1){04}}
\put(55,10){\line(-1,-1){04}} \put(60,10){\line(-1,-1){04}}
\put(65,10){\line(-1,-1){04}} \put(70,10){\line(-1,-1){04}}



\put(08,35){\oval(16,8)} \put(08,35){\oval(15,7)}


\put(0,27.5){\makebox(0,0)[bl]{Unmanned vehicle,}}
\put(0,24){\makebox(0,0)[bl]{measuring probe}}
\put(0,21){\makebox(0,0)[bl]{(sensors,}}
\put(0,18){\makebox(0,0)[bl]{transmitter unit)}}


\put(32,22){\line(1,0){18}}

\put(32,10){\line(0,1){12}} \put(50,10){\line(0,1){12}}
\put(32.5,10){\line(0,1){12}}\put(49.5,10){\line(0,1){12}}

\put(33.4,17){\makebox(0,0)[bl]{Receiving,}}
\put(33.4,14){\makebox(0,0)[bl]{collection,}}
\put(33.4,11){\makebox(0,0)[bl]{processing}}

\put(36,27){\makebox(0,0)[bl]{Ground point of data}}
\put(33.5,24){\makebox(0,0)[bl]{collection and processing}}


\put(59,18){\makebox(0,0)[bl]{Operator}}

\put(56,20.5){\circle{03}}

\put(56,19){\line(0,-1){05}} \put(53,17){\line(1,0){6}}
\put(56,14){\line(-1,-1){4}} \put(56,14){\line(1,-1){4}}


\put(17,34){\line(3,-1){15}} \put(32,29){\line(-1,-1){4}}
\put(28,25){\vector(3,-1){6}}


\put(27,35){\makebox(0,0)[bl]{Radio}}
\put(27,32){\makebox(0,0)[bl]{channel}}

\end{picture}
\end{center}

\section{Basic Hierarchical Morphological Model}

 The following hierarchical system model is considered
 (\cite{lev06},\cite{levagg11})
 (Fig. 2):

\begin{center}
\begin{picture}(74,60)
\put(07.6,00){\makebox(0,0)[bl]{Fig. 2.
 Hierarchical  system model \cite{levagg11}}}


\put(38,58){\circle*{2.8}}


\put(23.6,47){\makebox(0,0)[bl]{System hierarchy}}


\put(01,36){\makebox(0,0)[bl]{System compo-}}
\put(01,33){\makebox(0,0)[bl]{nents (leaf nodes)}}

\put(04,33){\line(0,-1){8}}

\put(05,33){\line(2,-3){5.4}}

\put(6,33){\line(3,-1){23}}


\put(29.5,36){\makebox(0,0)[bl]{Sets of design}}
\put(29.5,33.5){\makebox(0,0)[bl]{alternatives}}

\put(33,33){\line(-1,-1){15}} \put(43,33){\line(1,-1){15}}


\put(52,36){\makebox(0,0)[bl]{Compatibility}}
\put(52,33){\makebox(0,0)[bl]{among design}}
\put(52,30.5){\makebox(0,0)[bl]{alternatives}}

\put(60.4,30){\line(-1,-1){20.5}}


\put(00,23){\line(1,0){74}}

\put(00,39){\line(2,1){37}} \put(74,39){\line(-2,1){37}}

\put(00,23){\line(0,1){16}} \put(74,23){\line(0,1){16}}

\put(02,24.5){\makebox(0,0)[bl]{\(1\)}}

\put(03,23){\circle*{1.8}} \put(03,18){\oval(6,8)}

\put(12,24.5){\makebox(0,0)[bl]{\(2\)}}

\put(13,23){\circle*{1.8}} \put(13,18){\oval(6,8)}

\put(29.5,24){\makebox(0,0)[bl]{\(\tau-1\)}}

\put(32,23){\circle*{1.8}} \put(32,18){\oval(6,8)}

\put(41,24.5){\makebox(0,0)[bl]{\(\tau\)}}

\put(42,23){\circle*{1.8}} \put(42,18){\oval(6,8)}

\put(57,24.5){\makebox(0,0)[bl]{\(m-1\)}}

\put(61,23){\circle*{1.8}} \put(61,18){\oval(6,8)}

\put(70,24.5){\makebox(0,0)[bl]{\(m\)}}

\put(71,23){\circle*{1.8}} \put(71,18){\oval(6,8)}

\put(19.5,17.5){\makebox(0,0)[bl]{. . .}}
\put(48.5,17.5){\makebox(0,0)[bl]{. . .}}

\put(19.5,9){\makebox(0,0)[bl]{. . .}}
\put(48.5,9){\makebox(0,0)[bl]{. . .}}
\put(06,06){\line(1,0){04}} \put(06,13){\line(1,0){04}}
\put(06,06){\line(0,1){7}} \put(10,06){\line(0,1){7}}

\put(07,06){\line(0,1){7}} \put(08,06){\line(0,1){7}}
\put(09,06){\line(0,1){7}}

\put(35,06){\line(1,0){04}} \put(35,13){\line(1,0){04}}
\put(35,06){\line(0,1){7}} \put(39,06){\line(0,1){7}}

\put(36,06){\line(0,1){7}} \put(37,06){\line(0,1){7}}
\put(38,06){\line(0,1){7}}

\put(64,06){\line(1,0){04}} \put(64,13){\line(1,0){04}}
\put(64,06){\line(0,1){7}} \put(68,06){\line(0,1){7}}

\put(65,06){\line(0,1){7}} \put(66,06){\line(0,1){7}}
\put(67,06){\line(0,1){7}}


\end{picture}
\end{center}

 (i) tree-like system model,

 (ii) set of leaf nodes as basic system parts/components,

 (iii) sets of DAs for each leaf node,

 (iv) estimates of  DAs
  (e.g., ordinal priorities, interval multiset estimates);
   and

 (v) estimates of compatibility between DAs
 (ordinal estimates).



 Generally, the hierarchical structure of  a telemetry system
 is the following (Fig. 3):

 {\bf 0.} Telemetry system
 \( S =  A \star  R \star L\):

 {\bf 1.} On-Board equipment
 \( A =  D \star  E \star F\):

 {\it 1.1.} power supply \(D\),

 {\it 1.2.} sensor elements \(E\),

 {\it 1.3.} data processing \(F\).

  {\bf 2.} Radio channel \(R\).

  {\bf 3.} Ground point \( L =  B \star  O\):

  {\it 3.1.} ground equipment  \( B = U \star V\):

 {\it 3.1.1.} power supply \(U\),

 {\it 3.1.2.} operator working place(s) \(V\),

 {\it 3.2.} operator(s) \(O\).

 \begin{center}
\begin{picture}(75,81)

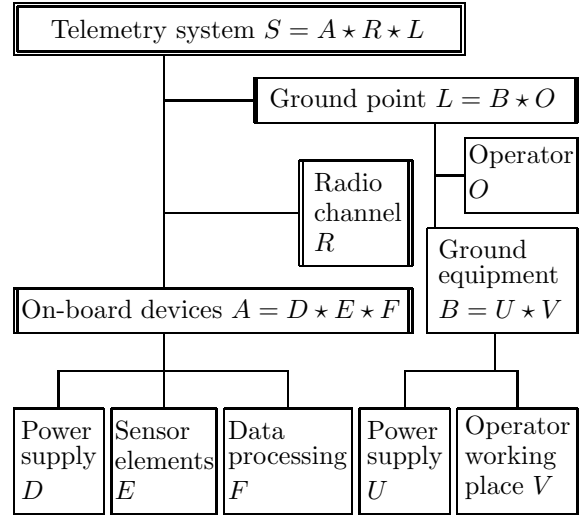
\put(01.5,02){\makebox(0,0)[bl]{Fig. 3. Structure of applied
 telemetry system}}

\put(00,10){\line(1,0){12}} \put(00,24){\line(1,0){12}}
\put(00,10){\line(0,1){14}} \put(12,10){\line(0,1){14}}

\put(01,20){\makebox(0,0)[bl]{Power}}
\put(01,16){\makebox(0,0)[bl]{supply}}
\put(01,12){\makebox(0,0)[bl]{\(D\)}}


\put(13,10){\line(1,0){14}} \put(13,24){\line(1,0){14}}
\put(13,10){\line(0,1){14}} \put(27,10){\line(0,1){14}}

\put(13.5,20){\makebox(0,0)[bl]{Sensor }}
\put(13.5,16){\makebox(0,0)[bl]{elements}}
\put(13.5,12){\makebox(0,0)[bl]{\(E\)}}


\put(28,10){\line(1,0){17}} \put(28,24){\line(1,0){17}}
\put(28,10){\line(0,1){14}} \put(45,10){\line(0,1){14}}

\put(28.5,20){\makebox(0,0)[bl]{Data }}
\put(28.5,16){\makebox(0,0)[bl]{processing}}
\put(28.5,12){\makebox(0,0)[bl]{\(F\)}}


\put(38,43){\line(1,0){15}} \put(38,57){\line(1,0){15}}
\put(38,43){\line(0,1){14}} \put(53,43){\line(0,1){14}}
\put(38.5,43){\line(0,1){14}} \put(52.5,43){\line(0,1){14}}

\put(40,53){\makebox(0,0)[bl]{Radio}}
\put(40,49){\makebox(0,0)[bl]{channel }}
\put(40,45){\makebox(0,0)[bl]{\(R\)}}

\put(38,50){\line(-1,0){18}}


\put(59,10){\line(1,0){16}} \put(59,24){\line(1,0){16}}
\put(59,10){\line(0,1){14}} \put(75,10){\line(0,1){14}}

\put(60,20){\makebox(0,0)[bl]{Operator}}
\put(60,16){\makebox(0,0)[bl]{working}}
\put(60,12){\makebox(0,0)[bl]{place \(V\)}}


\put(46,10){\line(1,0){12}} \put(46,24){\line(1,0){12}}
\put(46,10){\line(0,1){14}} \put(58,10){\line(0,1){14}}

\put(47,20){\makebox(0,0)[bl]{Power}}
\put(47,16){\makebox(0,0)[bl]{supply}}
\put(47,12){\makebox(0,0)[bl]{\(U\)}}


\put(55,34){\line(1,0){20}} \put(55,48){\line(1,0){20}}
\put(55,34){\line(0,1){14}} \put(75,34){\line(0,1){14}}

\put(56.5,44){\makebox(0,0)[bl]{Ground }}
\put(56.5,40){\makebox(0,0)[bl]{equipment}}
\put(56.5,36){\makebox(0,0)[bl]{\(B = U \star V\)}}


\put(64,34){\line(0,-1){05}}

\put(67,29){\line(0,-1){05}}

\put(67,29){\line(-1,0){15}}

\put(52,24){\line(0,1){5}}


\put(32,65){\line(-1,0){12}}

\put(32,62){\line(1,0){43}} \put(32,68){\line(1,0){43}}
\put(32,62){\line(0,1){06}} \put(75,62){\line(0,1){06}}
\put(32.4,62){\line(0,1){06}} \put(74.6,62){\line(0,1){06}}

\put(34,63.5){\makebox(0,0)[bl]{Ground point \(L = B \star O\)}}

\put(56,62){\line(0,-1){14}}


\put(60,55){\line(-1,0){4}}

\put(60,50){\line(1,0){15}} \put(60,60){\line(1,0){15}}
\put(60,50){\line(0,1){10}} \put(75,50){\line(0,1){10}}

\put(60.5,56){\makebox(0,0)[bl]{Operator}}
\put(60.5,52){\makebox(0,0)[bl]{\(O\)}}


\put(00,34){\line(1,0){53}} \put(00,40){\line(1,0){53}}
\put(00,34){\line(0,1){06}} \put(53,34){\line(0,1){06}}
\put(0.5,34){\line(0,1){06}} \put(52.5,34){\line(0,1){06}}

\put(01,36){\makebox(0,0)[bl]{On-board devices
 \(A = D \star E \star F\)}}

\put(20,34){\line(0,-1){05}}

\put(06,29){\line(0,-1){05}}

\put(06,29){\line(1,0){30.5}}

\put(20,29){\line(0,-1){05}}

\put(36.5,29){\line(0,-1){05}}


\put(0.5,71.5){\line(1,0){59}} \put(0.5,77.5){\line(1,0){59}}
\put(0.5,71.5){\line(0,1){06}} \put(59.5,71.5){\line(0,1){06}}

\put(00,71){\line(1,0){60}} \put(00,78){\line(1,0){60}}
\put(00,71){\line(0,1){07}} \put(60,71){\line(0,1){07}}

\put(05,72.65){\makebox(0,0)[bl]{Telemetry system \(S = A \star R
\star L\)}}

\put(20,71){\line(0,-1){31}}



\end{picture}
\end{center}

\section{Combinatorial Synthesis with Interval Multiset Estimates}

\subsection{Interval Multiset Estimates}

 Interval multiset estimates have been suggested by M.Sh. Levin
 in \cite{lev12a}.
 The approach consists in assignment of elements (\(1,2,3,...\))
 into an ordinal scale \([1,2,...,l]\).
 As a result, a multi-set based estimate is obtained,
 where a basis set involves all levels of the ordinal scale:
 \(\Omega = \{ 1,2,...,l\}\) (the levels are linear ordered:
 \(1 \succ 2 \succ 3 \succ ...\)) and
 the assessment problem (for each alternative)
 consists in selection of a multiset over set \(\Omega\) while taking into
 account two conditions:

 {\it 1.} cardinality of the selected multiset equals a specified
 number of elements \( \eta = 1,2,3,...\)
 (i.e., multisets of cardinality \(\eta \) are considered);

 {\it 2.} ``configuration'' of the multiset is the following:
 the selected elements of \(\Omega\) cover an interval over scale \([1,l]\)
 (i.e., ``interval multiset estimate'').

 Thus, an estimate \(e\) for an alternative \(A\) is
 (scale \([1,l]\), position-based form or position form):
 \(e(A) = (\eta_{1},...,\eta_{\iota},...,\eta_{l})\),
 where \(\eta_{\iota}\) corresponds to the number of elements at the
 level \(\iota\) (\(\iota = \overline{1,l}\)), or
 \(e(A) = \{ \overbrace{1,...,1}^{\eta_{1}},\overbrace{2,...2}^{\eta_{2}},
 \overbrace{3,...,3}^{\eta_{3}},...,\overbrace{l,...,l}^{\eta_{l}}
 \}\).
 The number of multisets of cardinality \(\eta\),
 with elements taken from a finite set of cardinality \(l\),
 is called the
 ``multiset coefficient'' or ``multiset number''
  (\cite{knuth98},\cite{yager86}):
 ~~\( \mu^{l,\eta} =
   \frac{l(l+1)(l+2)... (l+\eta-1) } {\eta!}
   \).
 This number corresponds to possible estimates
 (without taking into account interval condition 2).
 In the case of condition 2,
 the number of estimates is decreased.
 Generally, assessment problems based on interval multiset estimates
 can be denoted as follows: ~\(P^{l,\eta}\).
 A poset-like scale of interval multiset estimates for assessment problem \(P^{4,3}\)
 is presented
 in Fig. 4.
 The assessment problem will be used in our applied numerical
 examples.

 In addition, operations over multiset estimates
 are used \cite{lev12a}:
 integration, vector-like proximity, aggregation, and alignment.


 Integration of estimates (mainly, for composite systems)
 is based on summarization of the estimates by components (i.e.,
 positions).
 Let us consider \(n\) estimates (position form):~~
 estimate \(e^{1} = (\eta^{1}_{1},...,\eta^{1}_{\iota},...,\eta^{1}_{l})
 \),
  {\bf . . .},
 estimate \(e^{\kappa} = (\eta^{\kappa}_{1},...,\eta^{\kappa}_{\iota},...,\eta^{\kappa}_{l})
 \),
  {\bf . . .},
 estimate \(e^{n} = (\eta^{n}_{1},...,\eta^{n}_{\iota},...,\eta^{n}_{l})
 \).
 Then, the integrated estimate is:~
 estimate \(e^{I} = (\eta^{I}_{1},...,\eta^{I}_{\iota},...,\eta^{I}_{l})
 \),
 where
 \(\eta^{I}_{\iota} = \sum_{\kappa=1}^{n} \eta^{\kappa}_{\iota} ~~ \forall
 \iota = \overline{1,l}\).
 In fact, the operation \(\biguplus\) is used for multiset estimates:
 \(e^{I} = e^{1} \biguplus ... \biguplus e^{\kappa} \biguplus ... \biguplus e^{n}\).


 Further, vector-like proximity is described.
  Let \(A_{1}\) and \(A_{2}\) be two alternatives
 with corresponding
 interval multiset estimates
 \(e(A_{1})\), \(e(A_{2})\).
  Vector-like proximity for the alternatives above is:
 ~~\(\delta ( e(A_{1}), e(A_{2})) = (\delta^{-}(A_{1},A_{2}),\delta^{+}(A_{1},A_{2}))\),
 where vector components are:
 (i) \(\delta^{-}\) is the number of one-step changes:
 element of quality \(\iota + 1\) into element of quality \(\iota\) (\(\iota = \overline{1,l-1}\))
 (this corresponds to ``improvement'');
 (ii) \(\delta^{+}\) is the number of one-step changes:
 element of quality \(\iota\) into element of quality  \(\iota+1\) (\(\iota = \overline{1,l-1}\))
 (this corresponds to ``degradation'').
 It is assumed:
 ~\( | \delta ( e(A_{1}), e(A_{2})) | = | \delta^{-}(A_{1},A_{2}) | + |\delta^{+}(A_{1},A_{2})|
 \).

\begin{center}
\begin{picture}(68,140)

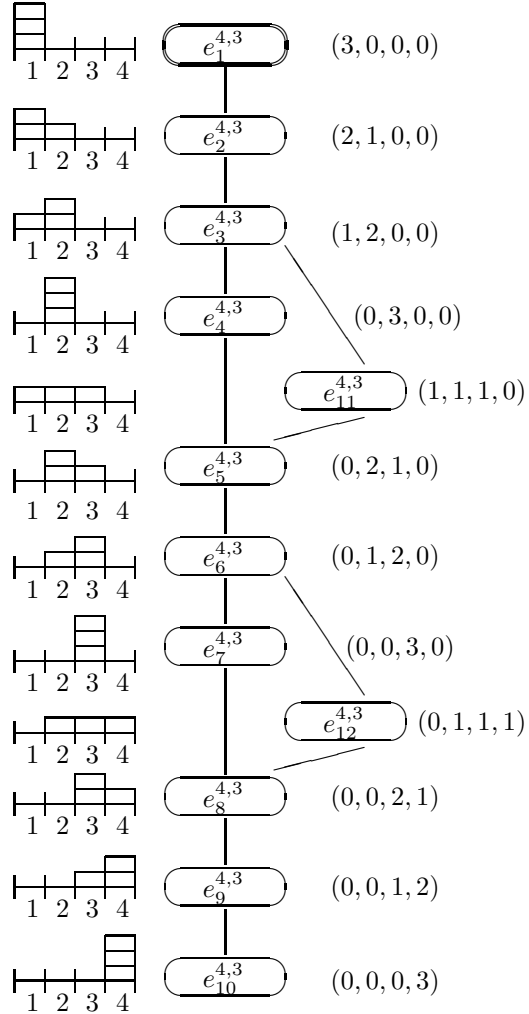
\put(08,00){\makebox(0,0)[bl] {Fig. 4. Scale, estimates
 (\(P^{4,3}\))}}




\put(25,130.7){\makebox(0,0)[bl]{\(e^{4,3}_{1}\) }}

\put(28,133){\oval(16,5)} \put(28,133){\oval(16.5,5.5)}

\put(42,131){\makebox(0,0)[bl]{\((3,0,0,0)\) }}

\put(00,132.5){\line(0,1){06}} \put(04,132.5){\line(0,1){06}}

\put(00,134.5){\line(1,0){4}} \put(00,136.5){\line(1,0){4}}
\put(00,138.5){\line(1,0){4}}


\put(00,132.5){\line(1,0){16}}

\put(00,131){\line(0,1){3}} \put(04,131){\line(0,1){3}}
\put(08,131){\line(0,1){3}} \put(12,131){\line(0,1){3}}
\put(16,131){\line(0,1){3}}

\put(01.5,128.5){\makebox(0,0)[bl]{\(1\)}}
\put(05.5,128.5){\makebox(0,0)[bl]{\(2\)}}
\put(09.5,128.5){\makebox(0,0)[bl]{\(3\)}}
\put(13.5,128.5){\makebox(0,0)[bl]{\(4\)}}


\put(28,124){\line(0,1){6}}


\put(25,118.7){\makebox(0,0)[bl]{\(e^{4,3}_{2}\) }}

\put(28,121){\oval(16,5)}

\put(42,119){\makebox(0,0)[bl]{\((2,1,0,0)\) }}

\put(00,120.5){\line(0,1){04}} \put(04,120.5){\line(0,1){04}}
\put(08,120.5){\line(0,1){02}}

\put(00,122.5){\line(1,0){8}} \put(00,124.5){\line(1,0){4}}

\put(00,120.5){\line(1,0){16}}

\put(00,119){\line(0,1){3}} \put(04,119){\line(0,1){3}}
\put(08,119){\line(0,1){3}} \put(12,119){\line(0,1){3}}
\put(16,119){\line(0,1){3}}

\put(01.5,116.5){\makebox(0,0)[bl]{\(1\)}}
\put(05.5,116.5){\makebox(0,0)[bl]{\(2\)}}
\put(09.5,116.5){\makebox(0,0)[bl]{\(3\)}}
\put(13.5,116.5){\makebox(0,0)[bl]{\(4\)}}


\put(28,112){\line(0,1){6}}


\put(25,106.7){\makebox(0,0)[bl]{\(e^{4,3}_{3}\) }}

\put(28,109){\oval(16,5)}

\put(42,106){\makebox(0,0)[bl]{\((1,2,0,0)\) }}

\put(00,108.5){\line(0,1){02}}

\put(04,108.5){\line(0,1){04}} \put(08,108.5){\line(0,1){04}}

\put(00,110.5){\line(1,0){8}} \put(04,112.5){\line(1,0){4}}

\put(00,108.5){\line(1,0){16}}

\put(00,107){\line(0,1){3}} \put(04,107){\line(0,1){3}}
\put(08,107){\line(0,1){3}} \put(12,107){\line(0,1){3}}
\put(16,107){\line(0,1){3}}

\put(01.5,104.5){\makebox(0,0)[bl]{\(1\)}}
\put(05.5,104.5){\makebox(0,0)[bl]{\(2\)}}
\put(09.5,104.5){\makebox(0,0)[bl]{\(3\)}}
\put(13.5,104.5){\makebox(0,0)[bl]{\(4\)}}


\put(28,100){\line(0,1){6}}


\put(25,94.7){\makebox(0,0)[bl]{\(e^{4,3}_{4}\) }}

\put(28,97){\oval(16,5)}

\put(45,95){\makebox(0,0)[bl]{\((0,3,0,0)\) }}


\put(04,96){\line(0,1){06}} \put(08,96){\line(0,1){06}}

\put(04,98){\line(1,0){4}} \put(04,100){\line(1,0){4}}
\put(04,102){\line(1,0){4}}

\put(00,96){\line(1,0){16}}

\put(00,94.5){\line(0,1){3}} \put(04,94.5){\line(0,1){3}}
\put(08,94.5){\line(0,1){3}} \put(12,94.5){\line(0,1){3}}
\put(16,94.5){\line(0,1){3}}

\put(01.5,92){\makebox(0,0)[bl]{\(1\)}}
\put(05.5,92){\makebox(0,0)[bl]{\(2\)}}
\put(09.5,92){\makebox(0,0)[bl]{\(3\)}}
\put(13.5,92){\makebox(0,0)[bl]{\(4\)}}


\put(28,80){\line(0,1){14}}



\put(25,74.7){\makebox(0,0)[bl]{\(e^{4,3}_{5}\) }}

\put(28,77){\oval(16,5)}

\put(42,75){\makebox(0,0)[bl]{\((0,2,1,0)\) }}

\put(04,75){\line(0,1){04}} \put(08,75){\line(0,1){04}}
\put(12,75){\line(0,1){02}}

\put(04,77){\line(1,0){8}} \put(04,79){\line(1,0){4}}

\put(00,75){\line(1,0){16}}

\put(00,73.5){\line(0,1){3}} \put(04,73.5){\line(0,1){3}}
\put(08,73.5){\line(0,1){3}} \put(12,73.5){\line(0,1){3}}
\put(16,73.5){\line(0,1){3}}

\put(01.5,70){\makebox(0,0)[bl]{\(1\)}}
\put(05.5,70){\makebox(0,0)[bl]{\(2\)}}
\put(09.5,70){\makebox(0,0)[bl]{\(3\)}}
\put(13.5,70){\makebox(0,0)[bl]{\(4\)}}


\put(28,68){\line(0,1){6}}


\put(25,62.7){\makebox(0,0)[bl]{\(e^{4,3}_{6}\) }}

\put(28,65){\oval(16,5)}

\put(42,63){\makebox(0,0)[bl]{\((0,1,2,0)\) }}

\put(04,63.5){\line(0,1){02}} \put(08,63.5){\line(0,1){04}}
\put(12,63.5){\line(0,1){04}}

\put(04,65.5){\line(1,0){8}} \put(08,67.5){\line(1,0){4}}

\put(00,63.5){\line(1,0){16}}

\put(00,62){\line(0,1){3}} \put(04,62){\line(0,1){3}}
\put(08,62){\line(0,1){3}} \put(12,62){\line(0,1){3}}
\put(16,62){\line(0,1){3}}

\put(01.5,59.5){\makebox(0,0)[bl]{\(1\)}}
\put(05.5,59.5){\makebox(0,0)[bl]{\(2\)}}
\put(09.5,59.5){\makebox(0,0)[bl]{\(3\)}}
\put(13.5,59.5){\makebox(0,0)[bl]{\(4\)}}


\put(28,56){\line(0,1){6}}

\put(25,50.7){\makebox(0,0)[bl]{\(e^{4,3}_{7}\) }}

\put(28,53){\oval(16,5)}

\put(44,51){\makebox(0,0)[bl]{\((0,0,3,0)\) }}

\put(08,51){\line(0,1){06}} \put(12,51){\line(0,1){06}}

\put(08,53){\line(1,0){4}} \put(08,55){\line(1,0){4}}
\put(08,57){\line(1,0){4}}

\put(00,51){\line(1,0){16}}

\put(00,49.5){\line(0,1){3}} \put(04,49.5){\line(0,1){3}}
\put(08,49.5){\line(0,1){3}} \put(12,49.5){\line(0,1){3}}
\put(16,49.5){\line(0,1){3}}

\put(01.5,47){\makebox(0,0)[bl]{\(1\)}}
\put(05.5,47){\makebox(0,0)[bl]{\(2\)}}
\put(09.5,47){\makebox(0,0)[bl]{\(3\)}}
\put(13.5,47){\makebox(0,0)[bl]{\(4\)}}


\put(28,36){\line(0,1){14}}

\put(25,30.7){\makebox(0,0)[bl]{\(e^{4,3}_{8}\) }}

\put(28,33){\oval(16,5)}

\put(42,31){\makebox(0,0)[bl]{\((0,0,2,1)\) }}

\put(08,32){\line(0,1){04}}

\put(12,32){\line(0,1){04}} \put(16,32){\line(0,1){02}}
\put(08,34){\line(1,0){8}} \put(08,36){\line(1,0){4}}

\put(00,32){\line(1,0){16}}

\put(00,30.5){\line(0,1){3}} \put(04,30.5){\line(0,1){3}}
\put(08,30.5){\line(0,1){3}} \put(12,30.5){\line(0,1){3}}
\put(16,30.5){\line(0,1){3}}

\put(01.5,28){\makebox(0,0)[bl]{\(1\)}}
\put(05.5,28){\makebox(0,0)[bl]{\(2\)}}
\put(09.5,28){\makebox(0,0)[bl]{\(3\)}}
\put(13.5,28){\makebox(0,0)[bl]{\(4\)}}


\put(28,24){\line(0,1){6}}

\put(25,18.7){\makebox(0,0)[bl]{\(e^{4,3}_{9}\) }}

\put(28,21){\oval(16,5)}

\put(42,19){\makebox(0,0)[bl]{\((0,0,1,2)\) }}

\put(08,21){\line(0,1){02}}

\put(12,21){\line(0,1){04}} \put(16,21){\line(0,1){04}}

\put(08,23){\line(1,0){8}} \put(12,25){\line(1,0){4}}

\put(00,21){\line(1,0){16}}

\put(00,19.5){\line(0,1){3}} \put(04,19.5){\line(0,1){3}}
\put(08,19.5){\line(0,1){3}} \put(12,19.5){\line(0,1){3}}
\put(16,19.5){\line(0,1){3}}

\put(01.5,17){\makebox(0,0)[bl]{\(1\)}}
\put(05.5,17){\makebox(0,0)[bl]{\(2\)}}
\put(09.5,17){\makebox(0,0)[bl]{\(3\)}}
\put(13.5,17){\makebox(0,0)[bl]{\(4\)}}


\put(28,12){\line(0,1){6}}


\put(25,06.7){\makebox(0,0)[bl]{\(e^{4,3}_{10}\) }}

\put(28,09){\oval(16,5)}

\put(42,6.5){\makebox(0,0)[bl]{\((0,0,0,3)\) }}

\put(12,8.5){\line(0,1){06}} \put(16,8.5){\line(0,1){06}}
\put(12,10.5){\line(1,0){4}} \put(12,12.5){\line(1,0){4}}
\put(12,14.5){\line(1,0){4}}

\put(00,8.5){\line(1,0){16}}

\put(00,07){\line(0,1){3}} \put(04,07){\line(0,1){3}}
\put(08,07){\line(0,1){3}} \put(12,07){\line(0,1){3}}
\put(16,07){\line(0,1){3}}

\put(01.5,4.5){\makebox(0,0)[bl]{\(1\)}}
\put(05.5,4.5){\makebox(0,0)[bl]{\(2\)}}
\put(09.5,4.5){\makebox(0,0)[bl]{\(3\)}}
\put(13.5,4.5){\makebox(0,0)[bl]{\(4\)}}


\put(46.5,90.5){\line(-2,3){10.5}}

\put(46.5,83.5){\line(-4,-1){12}}


\put(41,84.7){\makebox(0,0)[bl]{\(e^{4,3}_{11}\) }}

\put(44,87){\oval(16,5)}

\put(53.5,85){\makebox(0,0)[bl]{\((1,1,1,0)\) }}

\put(00,85.5){\line(0,1){02}}

\put(04,85.5){\line(0,1){02}} \put(08,85.5){\line(0,1){02}}
\put(12,85.5){\line(0,1){02}}

\put(00,87.5){\line(1,0){12}}

\put(00,85.5){\line(1,0){16}}

\put(00,84){\line(0,1){3}} \put(04,84){\line(0,1){3}}
\put(08,84){\line(0,1){3}} \put(12,84){\line(0,1){3}}
\put(16,84){\line(0,1){3}}

\put(01.5,81.5){\makebox(0,0)[bl]{\(1\)}}
\put(05.5,81.5){\makebox(0,0)[bl]{\(2\)}}
\put(09.5,81.5){\makebox(0,0)[bl]{\(3\)}}
\put(13.5,81.5){\makebox(0,0)[bl]{\(4\)}}


\put(46.5,46.5){\line(-2,3){10.5}}



\put(46.5,39.5){\line(-4,-1){12}}



\put(41,40.7){\makebox(0,0)[bl]{\(e^{4,3}_{12}\) }}

\put(44,43){\oval(16,5)}

\put(53.5,41){\makebox(0,0)[bl]{\((0,1,1,1)\) }}

\put(04,41.5){\line(0,1){02}}

\put(08,41.5){\line(0,1){02}} \put(12,41.5){\line(0,1){02}}
\put(16,41.5){\line(0,1){02}}

\put(04,43.5){\line(1,0){12}}

\put(00,41.5){\line(1,0){16}}

\put(00,40){\line(0,1){3}} \put(04,40){\line(0,1){3}}
\put(08,40){\line(0,1){3}} \put(12,40){\line(0,1){3}}
\put(16,40){\line(0,1){3}}

\put(01.5,37.5){\makebox(0,0)[bl]{\(1\)}}
\put(05.5,37.5){\makebox(0,0)[bl]{\(2\)}}
\put(09.5,37.5){\makebox(0,0)[bl]{\(3\)}}
\put(13.5,37.5){\makebox(0,0)[bl]{\(4\)}}


\end{picture}
\end{center}


 Now let us consider median estimates (aggregation)
  for the
 specified set of initial estimates
 (traditional approach).
 Let \(E = \{ e_{1},...,e_{\kappa},...,e_{n}\}\)
 be the set of specified estimates
 (or a corresponding set of specified alternatives),
 let \(\overline{D} \)
 be the set of all possible estimates
 (or a corresponding set of possible alternatives)
 (\( E  \subseteq \overline{D} \)).
%
  Thus, the median estimates
  (``generalized median'' \(M^{g}\) and ``set median'' \(M^{s}\)) are:
 ~~\(M^{g} =   \arg \min_{M \in \overline{D}}~
   \sum_{\kappa=1}^{n} ~  | \delta (M, e_{\kappa}) |; ~~
 M^{s} =   \arg \min_{M\in E} ~
   \sum_{\kappa=1}^{n} ~ | \delta (M, e_{\kappa}) |\).

\subsection{Morphological Design with Interval Multiset Estimates}

 A brief description  of combinatorial synthesis
 (HMMD) with ordinal estimates of design alternatives
 is the following
 (\cite{lev06},\cite{lev09},\cite{lev12morph},\cite{lev12a}).
 An examined composite
 (modular, decomposable) system consists
 of components and their interconnection or compatibility (IC).
 Basic assumptions of HMMD are the following:
 ~{\it (a)} a tree-like structure of the system;
 ~{\it (b)} a composite estimate for system quality
     that integrates components (subsystems, parts) qualities and
    qualities of IC (compatibility) across subsystems;
 ~{\it (c)} monotonic criteria for the system and its components;
 ~{\it (d)} quality of system components and IC are evaluated on the basis
    of coordinated ordinal scales.
 The designations are:
  ~(1) design alternatives (DAs) for leaf nodes of the model;
  ~(2) priorities of DAs (\(\iota = \overline{1,l}\);
      \(1\) corresponds to the best one);
  ~(3) ordinal compatibility for each pair of DAs
  (\(w=\overline{1,\nu}\); \(\nu\) corresponds to the best one).
 Let \(S\) be a system consisting of \(m\) parts (components):
 \(R(1),...,R(i),...,R(m)\).
 A set of design alternatives
 is generated for each system part above.
 The problem is:

~~

 {\it Find a composite design alternative}
 ~~ \(S=S(1)\star ...\star S(i)\star ...\star S(m)\)~~
 {\it of DAs (one representative design alternative}
 ~\(S(i)\)
 {\it for each system component/part}
  ~\(R(i)\), \(i=\overline{1,m}\)
  {\it )}
 {\it with non-zero compatibility}
 {\it between design alternatives.}

~~

 A discrete ``space'' of the system excellence
 (a poset)
 on the basis of the following vector is used:
 ~~\(N(S)=(w(S);e(S))\),
 ~where \(w(S)\) is the minimum of pairwise compatibility
 between DAs which correspond to different system components
 (i.e.,
 \(~\forall ~R_{j_{1}}\) and \( R_{j_{2}}\),
 \(1 \leq j_{1} \neq j_{2} \leq m\))
 in \(S\),
 ~\(e(S)=(\eta_{1},...,\eta_{\iota},...,\eta_{l})\),
 ~where \(\eta_{\iota}\) is the number of DAs of the \(\iota\)th quality in \(S\).
 Further,
  the problem is described as follows:
 \[ \max~ e(S),\]
  \[\max~ w(S),\]
 \[s.t.
  ~~~w(S) \geq 1  .\]
 As a result,
 we search for composite solutions
 which are nondominated by \(N(S)\)
 (i.e., Pareto-efficient).
 ``Maximization''  of \(e(S)\) is based on the corresponding poset.
 The considered combinatorial problem is NP-hard
 and an enumerative solving scheme is used.
%

%
 In the article, combinatorial synthesis is based on usage of multiset
 estimates of design alternatives for system parts.
 For the resultant system \(S = S(1) \star ... \star S(i) \star ... \star S(m) \)
 the same type of the multiset estimate is examined:
  an aggregated estimate (``generalized median'')
  of corresponding multiset estimates of its components
 (i.e., selected DAs).
 Thus, \( N(S) = (w(S);e(S))\), where
 \(e(S)\) is the ``generalized median'' of estimates of the solution
 components.
 Finally, the modified problem is:
 \[ \max~ e(S) = M^{g} =
  \arg \min_{M \in \overline{D} }~~
  \sum_{i=1}^{m} ~ |\delta (M, e(S_{i})) |,\]
 \[ \max~ w(S),\]
 \[s.t.
  ~~~w(S) \geq 1  .\]
 Here enumeration methods or heuristics are used
 (\cite{lev06},\cite{lev09},\cite{lev12morph},\cite{lev12a}).

\subsection{Multiple Choice Problem with Interval Multiset Estimates}

 Basic multiple choice problem is:
 (e.g., \cite{gar79}, \cite{keller04}):
 \[\max\sum_{i=1}^{m} \sum_{j=1}^{q_{i}} c_{ij} x_{ij}\]
 \[ s.t. ~~ \sum_{i=1}^{m} \sum_{j=1}^{q_{i}} a_{ij} x_{ij} \leq b,
  ~~ \sum_{j=1}^{q_{i}} x_{ij} \leq 1,~ i=\overline{1,m},\]
  \[ x_{ij} \in \{0, 1\}.\]

%
 In the case of multiset estimates of
 item ``utility'' \(e_{i}, i \in \{1,...,i,...,m\}\)
 (instead of \(c_{i}\)),
 the following aggregated multiset estimate can be used
 for the objective function  (``maximization'')
 \cite{lev12a}):
 (a) an aggregated multiset estimate as the ``generalized median'',
 (b) an aggregated multiset estimate as the ``set median'',
 and
 (c) an integrated  multiset estimate.

 A special case of multiple choice problem is considered:

 (1) multiset estimates of item ``utility''
  \(e_{i,j}, ~ i \in \{1,...,i,...,m\}, j = \overline{1,q_{i}}\)
 (instead of \(c_{ij}\)),

 (2) an aggregated multiset estimate as the ``generalized median''
 (or ``set median'')
 is used for the objective function (``maximization'').

 The  item set is:

   \(A= \bigcup_{i=1}^{m} A_{i}\),
 ~\( A_{i} = \{ (i,1),(i,2),...,(i,q_{i})\} \).

 Boolean variable \(x_{i,j}\) corresponds to selection of the
 item \((i,j)\).
 The solution  is a subset of the initial item set:
 \( S = \{ (i,j) | x_{i,j}=1 \} \).
  The problem is:
 \[ \max~ e(S) =   \max~ M = \]
 \[ \arg \min_{M \in \overline{D} }
 ~~
  \sum_{(i,j) \in S=\{(i,j)| x_{i,j}=1\}} ~ | \delta (M, e_{i,j}) |\]
 \[ s.t. ~ \sum_{i=1}^{m} \sum_{j=1}^{q_{i}}  a_{ij} x_{i,j} \leq b,
 ~ \sum_{j=1}^{q_{i}} x_{ij} =  1,
 ~ x_{ij} \in \{0, 1\}.\]
 Here the following algorithms can be used (as for basic multiple choice problem)
 (e.g., \cite{gar79},\cite{keller04},\cite{lev12a}):
 (i) enumerative methods including dynamic programming approach,
 (ii) heuristics (e.g,  greedy algorithms),
 (iii) approximation schemes
 (e.g., modifications of dynamic programming approach).
 Evidently, this problem is similar to the above-mentioned combinatorial synthesis
 problem without compatibility of the selected items (objects, alternatives).

\section{Example for On-Board Telemetry Subsystem}

 Here a numerical example for on-board telemetry
 subsystem is considered from \cite{levkhod07}.
 The initial morphological structure for on-board subsystem  is
 the following  (Fig. 5):

 {\bf 1.} On-board subsystem \(A = D \star E \star F\).

 {\it 1.1.} Power supply \(D = X \star Y \star Z \):
 {\it 1.1.1.} stabilizer \(X\):
 \(X_{1}\) (standard),
 \(X_{2}\) (transistorized),
 \(X_{3}\) (high-stability);
 {\it 1.1.2.} main source \(Y\):
 \(Y_{1}\) (Li-ion),
 \(Y_{2}\) (Cd-Mn),
 \(Y_{3}\) (Li);
 {\it 1.1.3.} emergency cell \(Z\):
 \(Z_{1}\) (Li-ion),
 \(Z_{2}\) (Cd-Mn),
 \(Z_{3}\) (Li).

 {\it 1.2.} Sensor elements \(E = I \star Q \star G \):
 {\it 1.2.1.} acceleration sensors  \(I\):
 \(I_{1}\) (ADXL),
 \(I_{2}\) (ADIS),
 \(I_{3}\) (MMA);
 {\it 1.2.2.} position sensors \(Q\):
 \(Q_{1}\) (SS12),
 \(Q_{2}\) (SS16),
 \(Q_{3}\) (SS19),
 \(Q_{4}\) (SS49),
 \(Q_{5}\) (SS59),
 \(Q_{6}\) (SS94);
 {\it 1.2.3.} global positioning system (GPS) \(G\):
 \(G_{1}\) (EB),
 \(G_{2}\) (GT),
 \(G_{3}\) (LS),
 \(G_{4}\) (ZX).

 {\it 1.3.} Data processing system  \(F = H \star C \star W \):
  {\it 1.3.1.} data storage unit \(H\):
 \(H_{1}\) (SRAM),
 \(H_{2}\) (DRAM),
 \(H_{3}\) (FRAM);
 {\it 1.3.2.} processing unit (CPU) \(C\):
 \(C_{1}\) (AVR),
 \(C_{2}\) (ARM),
 \(C_{3}\) (ADSP),
 \(C_{4}\) (BM);
 {\it 1.3.3.} data write unit \(W\):
 \(W_{1}\) (built-in ADC),
 \(W_{2}\) (external ADC I2C),
 \(W_{3}\) (external ADC SPI),
 \(W_{4}\) (external ADC 2W),
 \(W_{5}\) (external ADC UART(1)).

\begin{center}
\begin{picture}(69,148)

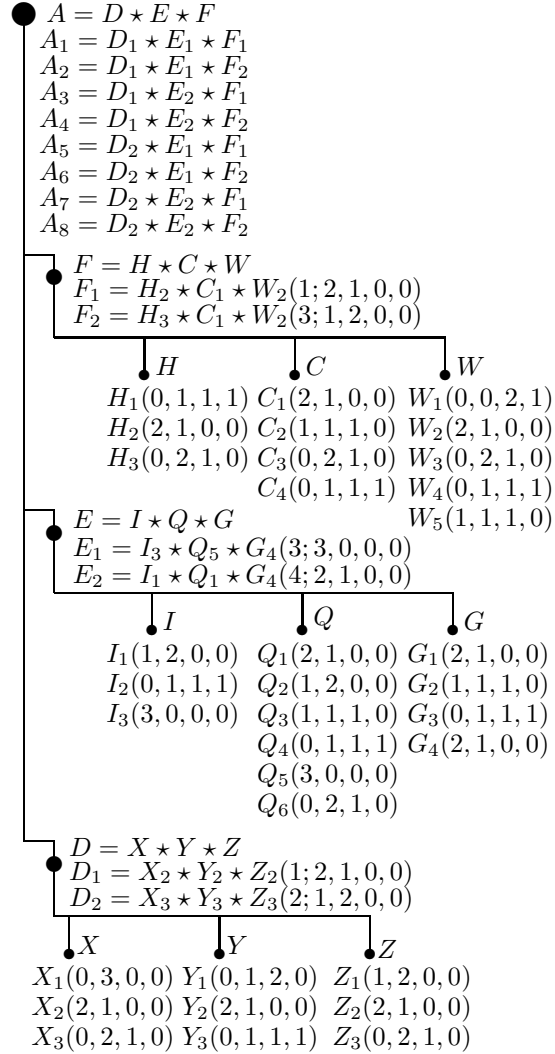
\put(05,00){\makebox(0,0)[bl] {Fig. 5. Structure on-board
subsystem}}

\put(00,144){\circle*{3}}

\put(00,34){\line(0,1){109}}

\put(04,101){\line(1,0){52}}

\put(04,67){\line(1,0){53}}

\put(00,34){\line(1,0){4}}



\put(03,143){\makebox(0,0)[bl]{\(A = D \star E \star F \)}}

\put(02,139){\makebox(0,0)[bl]{\(A_{1} = D_{1} \star E_{1}
 \star F_{1} \)}}

\put(02,135.5){\makebox(0,0)[bl]{\(A_{2} = D_{1} \star E_{1}
 \star F_{2} \)}}

\put(02,132){\makebox(0,0)[bl]{\(A_{3} = D_{1} \star E_{2}
 \star F_{1} \)}}

\put(02,128.5){\makebox(0,0)[bl]{\(A_{4} = D_{1} \star E_{2}
 \star F_{2}\)}}

\put(02,125){\makebox(0,0)[bl]{\(A_{5} = D_{2} \star E_{1}
 \star F_{1} \)}}

\put(02,121.5){\makebox(0,0)[bl]{\(A_{6} = D_{2} \star E_{1}
 \star F_{2} \)}}

\put(02,118){\makebox(0,0)[bl]{\(A_{7} = D_{2} \star E_{2}
 \star F_{1} \)}}

\put(02,114.5){\makebox(0,0)[bl]{\(A_{8} = D_{2} \star E_{2}
 \star F_{2} \)}}


\put(04,29){\line(0,1){05}}

\put(06,32){\makebox(0,0)[bl]{\(D=X \star Y \star Z \)}}

\put(06,28){\makebox(0,0)[bl]{\(D_{1}=X_{2} \star Y_{2} \star
 Z_{2} (1;2,1,0,0)\)}}

\put(06,24.5){\makebox(0,0)[bl]{\(D_{2}=X_{3} \star Y_{3} \star
 Z_{3} (2;1,2,0,0)\)}}

\put(04,31){\circle*{2}}


\put(04,24){\line(0,1){05}}


\put(04,24){\line(1,0){42}}


\put(06,19){\line(0,1){05}} \put(06,19){\circle*{1.5}}
\put(07,19){\makebox(0,0)[bl]{\(X\)}}
\put(01,14){\makebox(0,0)[bl]{\(X_{1}(0,3,0,0)\)}}
\put(01,10){\makebox(0,0)[bl]{\(X_{2}(2,1,0,0)\)}}
\put(01,06){\makebox(0,0)[bl]{\(X_{3}(0,2,1,0)\)}}


\put(26,19){\line(0,1){05}} \put(26,19){\circle*{1.5}}
\put(27,19){\makebox(0,0)[bl]{\(Y\)}}
\put(21,14){\makebox(0,0)[bl]{\(Y_{1}(0,1,2,0)\)}}
\put(21,10){\makebox(0,0)[bl]{\(Y_{2}(2,1,0,0)\)}}
\put(21,06){\makebox(0,0)[bl]{\(Y_{3}(0,1,1,1)\)}}


\put(46,19){\line(0,1){05}} \put(46,19){\circle*{1.5}}
\put(47,18.5){\makebox(0,0)[bl]{\(Z\)}}
\put(41,14){\makebox(0,0)[bl]{\(Z_{1}(1,2,0,0)\)}}
\put(41,10){\makebox(0,0)[bl]{\(Z_{2}(2,1,0,0)\)}}
\put(41,06){\makebox(0,0)[bl]{\(Z_{3}(0,2,1,0)\)}}


\put(00,78){\line(1,0){04}}

 \put(04,67){\line(0,1){11}}

\put(06.5,75){\makebox(0,0)[bl]{\(E= I \star Q \star G\)}}

\put(06.5,71){\makebox(0,0)[bl]{\(E_{1} = I_{3} \star Q_{5}
 \star G_{4} (3;3,0,0,0)\)}}

\put(06.5,67.5){\makebox(0,0)[bl]{\(E_{2} = I_{1} \star Q_{1}
 \star G_{4} (4;2,1,0,0)\)}}

\put(04,75){\circle*{2}}


\put(17,62){\line(0,1){05}}

\put(18.5,62){\makebox(0,0)[bl]{\(I\)}}

\put(17,62){\circle*{1.5}}

\put(11,57){\makebox(0,0)[bl]{\(I_{1}(1,2,0,0)\)}}
\put(11,53){\makebox(0,0)[bl]{\(I_{2}(0,1,1,1)\)}}
\put(11,49){\makebox(0,0)[bl]{\(I_{3}(3,0,0,0)\)}}


\put(37,62){\line(0,1){05}}

\put(38.5,62){\makebox(0,0)[bl]{\(Q\)}}

\put(37,62){\circle*{1.5}}

\put(31,57){\makebox(0,0)[bl]{\(Q_{1}(2,1,0,0)\)}}
\put(31,53){\makebox(0,0)[bl]{\(Q_{2}(1,2,0,0)\)}}
\put(31,49){\makebox(0,0)[bl]{\(Q_{3}(1,1,1,0)\)}}
\put(31,45){\makebox(0,0)[bl]{\(Q_{4}(0,1,1,1)\)}}
\put(31,41){\makebox(0,0)[bl]{\(Q_{5}(3,0,0,0)\)}}
\put(31,37){\makebox(0,0)[bl]{\(Q_{6}(0,2,1,0)\)}}


\put(57,62){\line(0,1){05}}

\put(58.5,62){\makebox(0,0)[bl]{\(G\)}}

\put(57,62){\circle*{1.5}}

\put(51,57){\makebox(0,0)[bl]{\(G_{1}(2,1,0,0)\)}}
\put(51,53){\makebox(0,0)[bl]{\(G_{2}(1,1,1,0)\)}}
\put(51,49){\makebox(0,0)[bl]{\(G_{3}(0,1,1,1)\)}}
\put(51,45){\makebox(0,0)[bl]{\(G_{4}(2,1,0,0)\)}}


\put(00,112){\line(1,0){04}}

 \put(04,101){\line(0,1){11}}

\put(06.5,109.5){\makebox(0,0)[bl]{\(F = H \star C \star W\)}}

\put(06.5,105.5){\makebox(0,0)[bl]{\(F_{1} = H_{2} \star C_{1}
\star W_{2} (1;2,1,0,0)\)}}

\put(06.5,102){\makebox(0,0)[bl]{\(F_{2} = H_{3} \star C_{1} \star
W_{2} (3;1,2,0,0)\)}}

\put(04,109){\circle*{2}}


\put(16,96){\line(0,1){05}}

\put(17.5,96){\makebox(0,0)[bl]{\(H\)}}

\put(16,96){\circle*{1.5}}

\put(11,91){\makebox(0,0)[bl]{\(H_{1}(0,1,1,1)\)}}
\put(11,87){\makebox(0,0)[bl]{\(H_{2}(2,1,0,0)\)}}
\put(11,83){\makebox(0,0)[bl]{\(H_{3}(0,2,1,0)\)}}


\put(36,96){\line(0,1){05}}

\put(37.5,96){\makebox(0,0)[bl]{\(C\)}}

\put(36,96){\circle*{1.5}}

\put(31,91){\makebox(0,0)[bl]{\(C_{1}(2,1,0,0)\)}}
\put(31,87){\makebox(0,0)[bl]{\(C_{2}(1,1,1,0)\)}}
\put(31,83){\makebox(0,0)[bl]{\(C_{3}(0,2,1,0)\)}}
\put(31,79){\makebox(0,0)[bl]{\(C_{4}(0,1,1,1)\)}}


\put(56,96){\line(0,1){05}}

\put(57.5,96){\makebox(0,0)[bl]{\(W\)}}

\put(56,96){\circle*{1.5}}

\put(51,91){\makebox(0,0)[bl]{\(W_{1}(0,0,2,1)\)}}
\put(51,87){\makebox(0,0)[bl]{\(W_{2}(2,1,0,0)\)}}
\put(51,83){\makebox(0,0)[bl]{\(W_{3}(0,2,1,0)\)}}
\put(51,79){\makebox(0,0)[bl]{\(W_{4}(0,1,1,1)\)}}
\put(51,75){\makebox(0,0)[bl]{\(W_{5}(1,1,1,0)\)}}

\end{picture}
\end{center}

 Interval multiset estimates of DAs
 are shown in Fig. 5 in parentheses
 (expert judgment).
 Ordinal estimates of compatibility are presented in
 Tables 1, 2, 3 (expert judgment, from \cite{levkhod07}).
%
 Note the initial combinatorial set
 of design solutions includes
  \(116640 \) possible system combinations
 (i.e.,
 \((3 \times 3 \times 3) \times
 (3 \times 6 \times 4) \times
 (3 \times 4 \times 5)\)).

\begin{center}
\begin{picture}(37,44)
\put(01,39){\makebox(0,0)[bl]{Table 1. Compatibility}}

\put(00,0){\line(1,0){37}} \put(00,31){\line(1,0){37}}
\put(00,37){\line(1,0){37}}

\put(00,0){\line(0,1){37}} \put(07,0){\line(0,1){37}}
\put(37,0){\line(0,1){37}}

\put(01,26.5){\makebox(0,0)[bl]{\(X_{1}\)}}
\put(01,21.5){\makebox(0,0)[bl]{\(X_{2}\)}}
\put(01,16.5){\makebox(0,0)[bl]{\(X_{3}\)}}
\put(01,11.5){\makebox(0,0)[bl]{\(Y_{1}\)}}
\put(01,06.5){\makebox(0,0)[bl]{\(Y_{2}\)}}
\put(01,01.5){\makebox(0,0)[bl]{\(Y_{3}\)}}

\put(12,31){\line(0,1){6}} \put(17,31){\line(0,1){6}}
\put(22,31){\line(0,1){6}} \put(27,31){\line(0,1){6}}
\put(32,31){\line(0,1){6}}

\put(07.7,32.4){\makebox(0,0)[bl]{\(Y_{1}\)}}
\put(12.7,32.4){\makebox(0,0)[bl]{\(Y_{2}\)}}
\put(17.7,32.4){\makebox(0,0)[bl]{\(Y_{3}\)}}
\put(22.7,32.4){\makebox(0,0)[bl]{\(Z_{1}\)}}
\put(27.7,32.4){\makebox(0,0)[bl]{\(Z_{2}\)}}
\put(32.7,32.4){\makebox(0,0)[bl]{\(Z_{3}\)}}

\put(09,27){\makebox(0,0)[bl]{\(3\)}}
\put(14,27){\makebox(0,0)[bl]{\(2\)}}
\put(19,27){\makebox(0,0)[bl]{\(2\)}}
\put(24,27){\makebox(0,0)[bl]{\(1\)}}
\put(29,27){\makebox(0,0)[bl]{\(1\)}}
\put(34,27){\makebox(0,0)[bl]{\(1\)}}

\put(09,22){\makebox(0,0)[bl]{\(2\)}}
\put(14,22){\makebox(0,0)[bl]{\(1\)}}
\put(19,22){\makebox(0,0)[bl]{\(1\)}}
\put(24,22){\makebox(0,0)[bl]{\(1\)}}
\put(29,22){\makebox(0,0)[bl]{\(1\)}}
\put(34,22){\makebox(0,0)[bl]{\(1\)}}

\put(09,17){\makebox(0,0)[bl]{\(4\)}}
\put(14,17){\makebox(0,0)[bl]{\(3\)}}
\put(19,17){\makebox(0,0)[bl]{\(3\)}}
\put(24,17){\makebox(0,0)[bl]{\(1\)}}
\put(29,17){\makebox(0,0)[bl]{\(1\)}}
\put(34,17){\makebox(0,0)[bl]{\(2\)}}

\put(24,12){\makebox(0,0)[bl]{\(2\)}}
\put(29,12){\makebox(0,0)[bl]{\(1\)}}
\put(34,12){\makebox(0,0)[bl]{\(1\)}}

\put(24,07){\makebox(0,0)[bl]{\(1\)}}
\put(29,07){\makebox(0,0)[bl]{\(2\)}}
\put(34,07){\makebox(0,0)[bl]{\(1\)}}

\put(24,02){\makebox(0,0)[bl]{\(1\)}}
\put(29,02){\makebox(0,0)[bl]{\(1\)}}
\put(34,02){\makebox(0,0)[bl]{\(2\)}}

\end{picture}
\end{center}

\begin{center}
\begin{picture}(57,59)
\put(011,54){\makebox(0,0)[bl]{Table 2. Compatibility}}

\put(00,0){\line(1,0){57}} \put(00,46){\line(1,0){57}}
\put(00,52){\line(1,0){57}}

\put(00,0){\line(0,1){52}} \put(07,0){\line(0,1){52}}
\put(57,0){\line(0,1){52}}

\put(01,41.5){\makebox(0,0)[bl]{\(I_{1}\)}}
\put(01,36.5){\makebox(0,0)[bl]{\(I_{2}\)}}
\put(01,31.5){\makebox(0,0)[bl]{\(I_{3}\)}}
\put(01,26.5){\makebox(0,0)[bl]{\(Q_{1}\)}}
\put(01,21.5){\makebox(0,0)[bl]{\(Q_{2}\)}}
\put(01,16.5){\makebox(0,0)[bl]{\(Q_{3}\)}}
\put(01,11.5){\makebox(0,0)[bl]{\(Q_{4}\)}}
\put(01,06.5){\makebox(0,0)[bl]{\(Q_{5}\)}}
\put(01,01.5){\makebox(0,0)[bl]{\(Q_{6}\)}}

\put(12,46){\line(0,1){6}} \put(17,46){\line(0,1){6}}
\put(22,46){\line(0,1){6}} \put(27,46){\line(0,1){6}}
\put(32,46){\line(0,1){6}} \put(37,46){\line(0,1){6}}
\put(42,46){\line(0,1){6}} \put(47,46){\line(0,1){6}}
\put(52,46){\line(0,1){6}}

\put(07.5,47.4){\makebox(0,0)[bl]{\(Q_{1}\)}}
\put(12.5,47.4){\makebox(0,0)[bl]{\(Q_{2}\)}}
\put(17.5,47.4){\makebox(0,0)[bl]{\(Q_{3}\)}}
\put(22.5,47.4){\makebox(0,0)[bl]{\(Q_{4}\)}}
\put(27.5,47.4){\makebox(0,0)[bl]{\(Q_{5}\)}}
\put(32.5,47.4){\makebox(0,0)[bl]{\(Q_{6}\)}}
\put(37.5,47.4){\makebox(0,0)[bl]{\(G_{1}\)}}
\put(42.5,47.4){\makebox(0,0)[bl]{\(G_{2}\)}}
\put(47.5,47.4){\makebox(0,0)[bl]{\(G_{3}\)}}
\put(52.5,47.4){\makebox(0,0)[bl]{\(G_{4}\)}}

\put(09,42){\makebox(0,0)[bl]{\(4\)}}
\put(14,42){\makebox(0,0)[bl]{\(4\)}}
\put(19,42){\makebox(0,0)[bl]{\(4\)}}
\put(24,42){\makebox(0,0)[bl]{\(4\)}}
\put(29,42){\makebox(0,0)[bl]{\(4\)}}
\put(34,42){\makebox(0,0)[bl]{\(4\)}}
\put(39,42){\makebox(0,0)[bl]{\(3\)}}
\put(44,42){\makebox(0,0)[bl]{\(4\)}}
\put(49,42){\makebox(0,0)[bl]{\(4\)}}
\put(54,42){\makebox(0,0)[bl]{\(4\)}}

\put(09,37){\makebox(0,0)[bl]{\(3\)}}
\put(14,37){\makebox(0,0)[bl]{\(3\)}}
\put(19,37){\makebox(0,0)[bl]{\(3\)}}
\put(24,37){\makebox(0,0)[bl]{\(3\)}}
\put(29,37){\makebox(0,0)[bl]{\(3\)}}
\put(34,37){\makebox(0,0)[bl]{\(3\)}}
\put(39,37){\makebox(0,0)[bl]{\(2\)}}
\put(44,37){\makebox(0,0)[bl]{\(3\)}}
\put(49,37){\makebox(0,0)[bl]{\(3\)}}
\put(54,37){\makebox(0,0)[bl]{\(1\)}}

\put(09,32){\makebox(0,0)[bl]{\(3\)}}
\put(14,32){\makebox(0,0)[bl]{\(3\)}}
\put(19,32){\makebox(0,0)[bl]{\(3\)}}
\put(24,32){\makebox(0,0)[bl]{\(3\)}}
\put(29,32){\makebox(0,0)[bl]{\(2\)}}
\put(34,32){\makebox(0,0)[bl]{\(2\)}}
\put(39,32){\makebox(0,0)[bl]{\(1\)}}
\put(44,32){\makebox(0,0)[bl]{\(1\)}}
\put(49,32){\makebox(0,0)[bl]{\(1\)}}
\put(54,32){\makebox(0,0)[bl]{\(3\)}}

\put(39,27){\makebox(0,0)[bl]{\(3\)}}
\put(44,27){\makebox(0,0)[bl]{\(2\)}}
\put(49,27){\makebox(0,0)[bl]{\(3\)}}
\put(54,27){\makebox(0,0)[bl]{\(4\)}}

\put(39,22){\makebox(0,0)[bl]{\(1\)}}
\put(44,22){\makebox(0,0)[bl]{\(1\)}}
\put(49,22){\makebox(0,0)[bl]{\(3\)}}
\put(54,22){\makebox(0,0)[bl]{\(1\)}}

\put(39,17){\makebox(0,0)[bl]{\(2\)}}
\put(44,17){\makebox(0,0)[bl]{\(2\)}}
\put(49,17){\makebox(0,0)[bl]{\(3\)}}
\put(54,17){\makebox(0,0)[bl]{\(4\)}}

\put(39,12){\makebox(0,0)[bl]{\(2\)}}
\put(44,12){\makebox(0,0)[bl]{\(2\)}}
\put(49,12){\makebox(0,0)[bl]{\(2\)}}
\put(54,12){\makebox(0,0)[bl]{\(4\)}}

\put(39,07){\makebox(0,0)[bl]{\(2\)}}
\put(44,07){\makebox(0,0)[bl]{\(2\)}}
\put(49,07){\makebox(0,0)[bl]{\(2\)}}
\put(54,07){\makebox(0,0)[bl]{\(4\)}}

\put(39,02){\makebox(0,0)[bl]{\(2\)}}
\put(44,02){\makebox(0,0)[bl]{\(2\)}}
\put(49,02){\makebox(0,0)[bl]{\(2\)}}
\put(54,02){\makebox(0,0)[bl]{\(4\)}}

\end{picture}
\end{center}

\begin{center}
\begin{picture}(61,49)
\put(13.5,44){\makebox(0,0)[bl]{Table 3. Compatibility}}

\put(00,0){\line(1,0){61}} \put(00,36){\line(1,0){61}}
\put(00,42){\line(1,0){61}}

\put(00,0){\line(0,1){42}} \put(07,0){\line(0,1){42}}
\put(61,0){\line(0,1){42}}

\put(01,31.5){\makebox(0,0)[bl]{\(H_{1}\)}}
\put(01,26.5){\makebox(0,0)[bl]{\(H_{2}\)}}
\put(01,21.5){\makebox(0,0)[bl]{\(H_{3}\)}}
\put(01,16.5){\makebox(0,0)[bl]{\(C_{1}\)}}
\put(01,11.5){\makebox(0,0)[bl]{\(C_{2}\)}}
\put(01,06.5){\makebox(0,0)[bl]{\(C_{3}\)}}
\put(01,01.5){\makebox(0,0)[bl]{\(C_{4}\)}}

\put(13,36){\line(0,1){6}} \put(19,36){\line(0,1){6}}
\put(25,36){\line(0,1){6}} \put(31,36){\line(0,1){6}}
\put(37,36){\line(0,1){6}} \put(43,36){\line(0,1){6}}
\put(49,36){\line(0,1){6}} \put(55,36){\line(0,1){6}}

\put(08,37.4){\makebox(0,0)[bl]{\(C_{1}\)}}
\put(14,37.4){\makebox(0,0)[bl]{\(C_{2}\)}}
\put(20,37.4){\makebox(0,0)[bl]{\(C_{3}\)}}
\put(26,37.4){\makebox(0,0)[bl]{\(C_{4}\)}}
\put(32,37.4){\makebox(0,0)[bl]{\(W_{1}\)}}
\put(38,37.4){\makebox(0,0)[bl]{\(W_{2}\)}}
\put(44,37.4){\makebox(0,0)[bl]{\(W_{3}\)}}
\put(50,37.4){\makebox(0,0)[bl]{\(W_{4}\)}}
\put(56,37.4){\makebox(0,0)[bl]{\(W_{5}\)}}


\put(09,32){\makebox(0,0)[bl]{\(3\)}}
\put(15,32){\makebox(0,0)[bl]{\(3\)}}
\put(21,32){\makebox(0,0)[bl]{\(3\)}}
\put(27,32){\makebox(0,0)[bl]{\(2\)}}
\put(33,32){\makebox(0,0)[bl]{\(3\)}}
\put(39,32){\makebox(0,0)[bl]{\(3\)}}
\put(45,32){\makebox(0,0)[bl]{\(3\)}}
\put(51,32){\makebox(0,0)[bl]{\(3\)}}
\put(57,32){\makebox(0,0)[bl]{\(3\)}}


\put(09,27){\makebox(0,0)[bl]{\(1\)}}
\put(15,27){\makebox(0,0)[bl]{\(1\)}}
\put(21,27){\makebox(0,0)[bl]{\(2\)}}
\put(27,27){\makebox(0,0)[bl]{\(3\)}}
\put(33,27){\makebox(0,0)[bl]{\(2\)}}
\put(39,27){\makebox(0,0)[bl]{\(3\)}}
\put(45,27){\makebox(0,0)[bl]{\(3\)}}
\put(51,27){\makebox(0,0)[bl]{\(2\)}}
\put(57,27){\makebox(0,0)[bl]{\(2\)}}


\put(09,22){\makebox(0,0)[bl]{\(4\)}}
\put(15,22){\makebox(0,0)[bl]{\(3\)}}
\put(21,22){\makebox(0,0)[bl]{\(3\)}}
\put(27,22){\makebox(0,0)[bl]{\(3\)}}
\put(33,22){\makebox(0,0)[bl]{\(3\)}}
\put(39,22){\makebox(0,0)[bl]{\(3\)}}
\put(45,22){\makebox(0,0)[bl]{\(3\)}}
\put(51,22){\makebox(0,0)[bl]{\(3\)}}
\put(57,22){\makebox(0,0)[bl]{\(3\)}}


\put(33,17){\makebox(0,0)[bl]{\(3\)}}
\put(39,17){\makebox(0,0)[bl]{\(3\)}}
\put(45,17){\makebox(0,0)[bl]{\(3\)}}
\put(51,17){\makebox(0,0)[bl]{\(3\)}}
\put(57,17){\makebox(0,0)[bl]{\(3\)}}


\put(33,12){\makebox(0,0)[bl]{\(3\)}}
\put(39,12){\makebox(0,0)[bl]{\(3\)}}
\put(45,12){\makebox(0,0)[bl]{\(3\)}}
\put(51,12){\makebox(0,0)[bl]{\(3\)}}
\put(57,12){\makebox(0,0)[bl]{\(3\)}}


\put(33,07){\makebox(0,0)[bl]{\(3\)}}
\put(39,07){\makebox(0,0)[bl]{\(3\)}}
\put(45,07){\makebox(0,0)[bl]{\(3\)}}
\put(51,07){\makebox(0,0)[bl]{\(3\)}}
\put(57,07){\makebox(0,0)[bl]{\(3\)}}


\put(33,02){\makebox(0,0)[bl]{\(1\)}}
\put(39,02){\makebox(0,0)[bl]{\(1\)}}
\put(45,02){\makebox(0,0)[bl]{\(1\)}}
\put(51,02){\makebox(0,0)[bl]{\(1\)}}
\put(57,02){\makebox(0,0)[bl]{\(1\)}}

\end{picture}
\end{center}

\subsection{Composite Solutions}

 The obtained Pareto-efficient composite DAs for
 composite components are the following:

 (1) for \(D\):~
 \(D_{1}=X_{2} \star Y_{2} \star Z_{2}\),
 \( N(D_{1}) =  (1;2,1,0,0)\);

 \(D_{2}=X_{3} \star Y_{3} \star Z_{3}\),
  \(N(D_{2}) = (2;1,2,0,0)\);

 (2) for \(E\):~
  \(E_{1} = I_{3} \star Q_{5} \star G_{4}\),
  \( N(E_{1}) = (3;3,0,0,0)\);

 \(E_{2} = I_{1} \star Q_{1} \star G_{4}\),
  \(N(E_{2}) = (4;2,1,0,0)\);

 (3) for \(F\):~
 \(F_{1} = H_{2} \star C_{1} \star W_{2}\),
  \(N(F_{1}) = (1;2,1,0,0)\);

 \(F_{2} = H_{3} \star C_{1} \star W_{2} \),
  \(N(F_{2}) = (3;1,2,0,0)\).

 Fig. 6 illustrates
 ``discrete space'' (poset, each component corresponds to Fig. 4)
 of quality for subsystem \(F\).

\begin{center}
\begin{picture}(69,62)
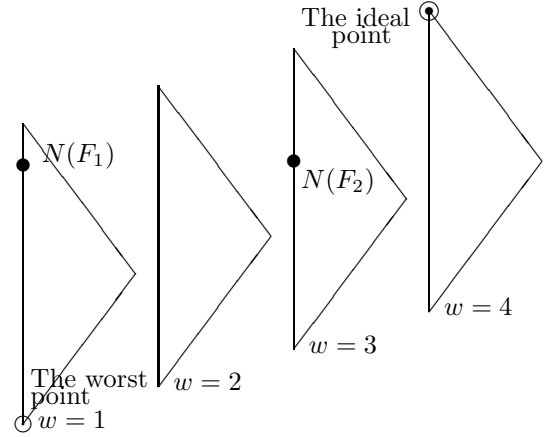
\put(06,0){\makebox(0,0)[bl]{Fig. 6. Illustration for quality of
\(F\)}}

\put(00,05){\line(0,1){40}} \put(00,05){\line(3,4){15}}
\put(00,45){\line(3,-4){15}}

\put(18,10){\line(0,1){40}} \put(18,10){\line(3,4){15}}
\put(18,50){\line(3,-4){15}}

\put(36,15){\line(0,1){40}} \put(36,15){\line(3,4){15}}
\put(36,55){\line(3,-4){15}}

\put(54,20){\line(0,1){40}} \put(54,20){\line(3,4){15}}
\put(54,60){\line(3,-4){15}}

\put(00,39.5){\circle*{1.7}}

\put(02.5,39){\makebox(0,0)[bl]{\(N(F_{1}) \)}}

\put(36,40){\circle*{1.7}}
\put(37,35.5){\makebox(0,0)[bl]{\(N(F_{2})\)}}


\put(54,60){\circle*{1}} \put(54,60){\circle{2.3}}

\put(37,58){\makebox(0,0)[bl]{The ideal}}
\put(41,55){\makebox(0,0)[bl]{point}}

\put(02,04.5){\makebox(0,0)[bl]{\(w=1\)}}
\put(20,09.5){\makebox(0,0)[bl]{\(w=2\)}}
\put(38,14.5){\makebox(0,0)[bl]{\(w=3\)}}
\put(56,19.5){\makebox(0,0)[bl]{\(w=4\)}}

\put(01,10){\makebox(0,0)[bl]{The worst}}
\put(01,07){\makebox(0,0)[bl]{point}}

\put(00,05){\circle*{0.5}} \put(00,05){\circle{2}}

\end{picture}
\end{center}

 For the resultant system, eight  obtained combinations of
 DAs for system parts are considered:

 \(A_{1} = D_{1} \star E_{1} \star F_{1} =
  ( X_{2} \star Y_{2} \star Z_{2} ) \star
  ( I_{3} \star Q_{5} \star G_{4} ) \star
  ( H_{2} \star C_{1} \star W_{2} )\),

 \(A_{2} = D_{1} \star E_{1} \star F_{2} =
  ( X_{2} \star Y_{2} \star Z_{2} ) \star
  ( I_{3} \star Q_{5} \star G_{4} ) \star
  ( H_{3} \star C_{1} \star W_{2} )\),

 \(A_{3} = D_{1} \star E_{2} \star F_{1} =
  ( X_{2} \star Y_{2} \star Z_{2} ) \star
  ( I_{1} \star Q_{1} \star G_{4} ) \star
  ( H_{2} \star C_{1} \star W_{2} )\),

 \(A_{4} = D_{1} \star E_{2} \star F_{2} =
  ( X_{2} \star Y_{2} \star Z_{2} ) \star
  ( I_{1} \star Q_{1} \star G_{4} ) \star
  ( H_{3} \star C_{1} \star W_{2} )\),

 \(A_{5} = D_{2} \star E_{1} \star F_{1} =
  ( X_{3} \star Y_{3} \star Z_{3} ) \star
  ( I_{3} \star Q_{5} \star G_{4} ) \star
  ( H_{2} \star C_{1} \star W_{2} )\),

 \(A_{6} = D_{2} \star E_{1} \star F_{2} =
  ( X_{3} \star Y_{3} \star Z_{3} ) \star
  ( I_{3} \star Q_{5} \star G_{4} ) \star
  ( H_{3} \star C_{1} \star W_{2} )\),

 \(A_{7} = D_{2} \star E_{2} \star F_{1} =
  ( X_{3} \star Y_{3} \star Z_{3} ) \star
  ( I_{1} \star Q_{1} \star G_{4} ) \star
  ( H_{2} \star C_{1} \star W_{2} )\),
  and

 \(A_{8} = D_{2} \star E_{2} \star F_{2}   =
  ( X_{3} \star Y_{3} \star Z_{3} ) \star
  ( I_{1} \star Q_{1} \star G_{4} ) \star
  ( H_{3} \star C_{1} \star W_{2} )\).

\subsection{Analysis and Improvement}

 System improvement process can be based on the
 following
 (\cite{lev06},\cite{lev10}):
 (i) improvement of a system component (element),
 (ii) improvement of compatibility between system components,
 (iii) change a system structure, e.g.,
 extension of the system by addition of system components/parts.
 On the other hand,
 aggregation of several initial system solutions into
 a resultant one can be considered as the improvement as well
 \cite{levagg11}.
%
 Here  system improvement (or reconfiguration) actions
 by elements and by compatibility are briefly presented.
 Subsystem \(F = H \star C \star W \)
 is examined as an example (Table 4).

\begin{center}
\begin{picture}(71,57)

\put(02.3,53){\makebox(0,0)[bl] {Table 4. Bottlenecks, improvement
actions}}

\put(00,00){\line(1,0){71}} \put(00,39){\line(1,0){71}}
\put(00,51){\line(1,0){71}}

\put(00,0){\line(0,1){51}} \put(20,0){\line(0,1){51}}

\put(35,0){\line(0,1){51}} \put(71,0){\line(0,1){51}}

\put(01,47){\makebox(0,0)[bl]{Composite}}
\put(01,44){\makebox(0,0)[bl]{DAs}}

\put(21,47){\makebox(0,0)[bl]{Bottle-}}
\put(21,44){\makebox(0,0)[bl]{necks}}
\put(21,40){\makebox(0,0)[bl]{DA/IC}}

\put(42.5,47){\makebox(0,0)[bl]{Improvement}}
\put(46.5,44){\makebox(0,0)[bl]{actions}}
\put(48.5,40){\makebox(0,0)[bl]{\(w/e\)}}


\put(1,34){\makebox(0,0)[bl]{\(F_{1}\)}}

\put(25,34){\makebox(0,0)[bl]{\(W_{2}\)}}

\put(36,33.5){\makebox(0,0)[bl]{\((2,1,0,0) \Rightarrow
(3,0,0,0)\)}}


\put(1,30){\makebox(0,0)[bl]{\(F_{1}\)}}

\put(25,30){\makebox(0,0)[bl]{\(C_{1}\)}}

\put(36,29.5){\makebox(0,0)[bl]{\((2,1,0,0) \Rightarrow
(3,0,0,0)\)}}


\put(1,26){\makebox(0,0)[bl]{\(F_{1}\)}}

\put(25,26){\makebox(0,0)[bl]{\(H_{2}\)}}

\put(36,25.5){\makebox(0,0)[bl]{\((2,1,0,0) \Rightarrow
(3,0,0,0)\)}}


\put(1,22){\makebox(0,0)[bl]{\(F_{1}\)}}

\put(20.5,21.6){\makebox(0,0)[bl]{\((H_{2},W_{2})\)}}

\put(48.6,22.4){\makebox(0,0)[bl]{\(1 \Rightarrow 3\)}}


\put(1,18){\makebox(0,0)[bl]{\(F_{2}\)}}

\put(20.5,17.6){\makebox(0,0)[bl]{\((H_{3},W_{2})\)}}

\put(48.6,18.4){\makebox(0,0)[bl]{\(3 \Rightarrow 4\)}}


\put(1,14){\makebox(0,0)[bl]{\(F_{2}\)}}

\put(20.5,13.6){\makebox(0,0)[bl]{\((C_{1},W_{2})\)}}

\put(48.6,14.4){\makebox(0,0)[bl]{\(3 \Rightarrow 4\)}}


\put(1,10){\makebox(0,0)[bl]{\(F_{2}\)}}

\put(25,10){\makebox(0,0)[bl]{\(W_{2}\)}}

\put(36,09.5){\makebox(0,0)[bl]{\((2,1,0,0) \Rightarrow
(3,0,0,0)\)}}


\put(1,06){\makebox(0,0)[bl]{\(F_{2}\)}}

\put(25,06){\makebox(0,0)[bl]{\(C_{1}\)}}

\put(36,05.5){\makebox(0,0)[bl]{\((2,1,0,0) \Rightarrow
(3,0,0,0)\)}}


\put(1,02){\makebox(0,0)[bl]{\(F_{2}\)}}

\put(25,02){\makebox(0,0)[bl]{\(H_{3}\)}}

\put(36,01.5){\makebox(0,0)[bl]{\((0,2,1,0) \Rightarrow
(3,0,0,0)\)}}

\end{picture}
\end{center}

 The following hypothetical improvement process (by elements) for \(F_{2}\)
 is examined (binary variables  \(\{y_{ij}\}\) are used):

 (1) two versions for element \(W_{2}\):
 \(y_{11}\) (none),
 \(y_{12}\) (\((2,1,0,0) \Rightarrow (3,0,0,0) \));

 (2) two versions for element \(C_{1}\):
 \(y_{21}\) (none),
 \(y_{22}\) (\((2,1,0,0) \Rightarrow (3,0,0,0) \));

 (3) five versions for element \(H_{3}\):
 \(y_{31}\) (none),
 \(y_{32}\) (\((0,2,1,0) \Rightarrow (0,3,0,0) \)),
 \(y_{33}\) (\((0,2,1,0) \Rightarrow (1,2,0,0) \)),
 \(y_{34}\) (\((0,2,1,0) \Rightarrow (2,1,0,0) \)),
 \(y_{35}\) (\((0,2,1,0) \Rightarrow (3,0,0,0) \)).

 Table 5 contains binary variables (\(y_{ij}\)),
 improvement actions and their estimates (illustrative, expert judgment).
%
 Thus, the improvement problem is:
 \[ \arg \min_{M \in \overline{D}}
 ~~
  \sum_{(i,j) \in S=\{(i,j)| y_{ij}=1\}} ~ | \delta (M, e_{ij}) |\]
 \[s.t. ~ \sum_{i=1}^{3} \sum_{j=1}^{q_{i}}  a_{ij} y_{ij} \leq b,
 ~ \sum_{j=1}^{q_{j}} y_{ij} =  1,
 ~ y_{ij} \in \{0, 1\}.\]

\begin{center}
\begin{picture}(67,76)
\put(11,74){\makebox(0,0)[bl]{Table 5. Improvement of \(F_{2}\)}}

\put(00,0){\line(1,0){67}} \put(00,62){\line(1,0){67}}
\put(00,72){\line(1,0){67}}

\put(00,0){\line(0,1){72}} \put(35,00){\line(0,1){72}}
\put(55,0){\line(0,1){72}} \put(67,00){\line(0,1){72}}

\put(01,67.7){\makebox(0,0)[bl]{Improvement}}
\put(01,64.5){\makebox(0,0)[bl]{actions}}

\put(36.5,68.2){\makebox(0,0)[bl]{Multiset  }}
\put(36.5,64){\makebox(0,0)[bl]{estimate \(e_{ij}\)}}

\put(56.5,68){\makebox(0,0)[bl]{Cost}}
\put(56.5,63.8){\makebox(0,0)[bl]{(\(a_{ij}\))}}


\put(01,58){\makebox(0,0)[bl]{\(y_{11}\)}}
\put(07,57.5){\makebox(0,0)[bl]{(\(W_{2}\), none)}}

\put(01,54){\makebox(0,0)[bl]{\(y_{12}\)}}
\put(07,53.5){\makebox(0,0)[bl]{(\(W_{2} \Rightarrow
W_{2}^{1}\),}}

\put(07,50){\makebox(0,0)[bl]{improvement \(1\))}}


\put(01,46){\makebox(0,0)[bl]{\(y_{21}\)}}
\put(07,45.5){\makebox(0,0)[bl]{(\(C_{1}\), none) }}

\put(01,42){\makebox(0,0)[bl]{\(y_{22}\)}}
\put(07,41.5){\makebox(0,0)[bl]{(\(C_{1} \Rightarrow
C_{1}^{1}\),}}

\put(07,38){\makebox(0,0)[bl]{improvement \(1\))}}


\put(01,34){\makebox(0,0)[bl]{\(y_{31}\)}}
\put(07,33.5){\makebox(0,0)[bl]{(\(H_{3}\), none)}}

\put(01,30){\makebox(0,0)[bl]{\(y_{32}\)}}
\put(07,29.5){\makebox(0,0)[bl]{(\(H_{3} \Rightarrow
H_{3}^{1}\),}}

\put(07,26){\makebox(0,0)[bl]{improvement \(1\)) }}

\put(01,22){\makebox(0,0)[bl]{\(y_{33}\)}}
\put(07,21.5){\makebox(0,0)[bl]{(\(H_{3}\Rightarrow H_{3}^{2}\),}}
\put(07,18){\makebox(0,0)[bl]{improvement \(2\))}}

\put(01,14){\makebox(0,0)[bl]{\(y_{34}\)}}
\put(07,13.5){\makebox(0,0)[bl]{(\(H_{3} \Rightarrow
H_{3}^{3}\),}}

\put(07,10){\makebox(0,0)[bl]{improvement \(3\))}}

\put(01,6){\makebox(0,0)[bl]{\(y_{35}\)}}
\put(07,5.5){\makebox(0,0)[bl]{(\(H_{3} \Rightarrow H_{3}^{4}\),}}

\put(07,2){\makebox(0,0)[bl]{improvement \(4\))}}



\put(38,57.2){\makebox(0,0)[bl]{\((2,1,0,0)\)}}

\put(59,58){\makebox(0,0)[bl]{\(0\)}}


\put(38,53.2){\makebox(0,0)[bl]{\((3,0,0,0)\)}}

\put(58,54){\makebox(0,0)[bl]{\(17\)}}


\put(38,45.2){\makebox(0,0)[bl]{\((2,1,0,0)\)}}

\put(59,46){\makebox(0,0)[bl]{\(0\)}}


\put(38,41.2){\makebox(0,0)[bl]{\((3,0,0,0)\)}}

\put(58,42){\makebox(0,0)[bl]{\(15\)}}


\put(38,33.2){\makebox(0,0)[bl]{\((0,2,1,0)\)}}

\put(59,34){\makebox(0,0)[bl]{\(0\)}}


\put(38,29.2){\makebox(0,0)[bl]{\((0,3,0,0)\)}}

\put(59,30){\makebox(0,0)[bl]{\(1\)}}


\put(38,21.2){\makebox(0,0)[bl]{\((1,2,0,0)\)}}

\put(59,22){\makebox(0,0)[bl]{\(7\)}}


\put(38,13.2){\makebox(0,0)[bl]{\((2,1,0,0)\)}}

\put(58,14){\makebox(0,0)[bl]{\(13\)}}


\put(38,5.2){\makebox(0,0)[bl]{\((3,0,0,0)\)}}

\put(58,06){\makebox(0,0)[bl]{\(22\)}}

\end{picture}
\end{center}

 Some examples of the improvement solutions are:

  (1) ~\(b=1\):~
 \(y_{11}=1\)  (\(W_{2}\), none),
 \(y_{21}=1\)  (\(C_{1}\), none),
  \(y_{32}=1\)   (\(H_{3} \), improvement \(1\));

 \( {F}_{2} \Rightarrow
 \widetilde{F_{2}^{1}} =
  H_{3}^{1} \star C_{1} \star W_{2} \),
   \(e(\widetilde{F_{2}^{1}}) = (2,1,0,0)\);

  (2) ~\(b=45\):~~
 \(y_{12}=1\) (\(W_{2}\), improvement \(1\)),
 \(y_{22}=1\) (\( C_{1}\), improvement \(1\)),
  \(y_{34}=1\) (\( Z_{1}\), improvement \(3\)), ~
 \( {F}_{2} \Rightarrow
 \widetilde{F_{2}^{2}} =
  H_{3}^{3} \star C_{1}^{1} \star W_{2}^{1} \),
   \(e(\widetilde{F_{2}^{2}}) = (3,0,0,0)\).

\subsection{Aggregation of Solutions}

 Aggregation procedures for hierarchical structures
 are presented in \cite{levagg11}.
 Here, a simplified approach to aggregation
 (extension of a ``system kernel'' based on multiple choice problem)
 is considered for the obtained eight solutions:

 \(A_{1} =
  ( X_{2} \star Y_{2} \star Z_{2} ) \star
  ( I_{3} \star Q_{5} \star G_{4} ) \star
  ( H_{2} \star C_{1} \star W_{2} )\),

 \(A_{2} =
  ( X_{2} \star Y_{2} \star Z_{2} ) \star
  ( I_{3} \star Q_{5} \star G_{4} ) \star
  ( H_{3} \star C_{1} \star W_{2} )\),

 \(A_{3} =
  ( X_{2} \star Y_{2} \star Z_{2} ) \star
  ( I_{1} \star Q_{1} \star G_{4} ) \star
  ( H_{2} \star C_{1} \star W_{2} )\),

 \(A_{4} =
  ( X_{2} \star Y_{2} \star Z_{2} ) \star
  ( I_{1} \star Q_{1} \star G_{4} ) \star
  ( H_{3} \star C_{1} \star W_{2} )\),

 \(A_{5} =
  ( X_{3} \star Y_{3} \star Z_{3} ) \star
  ( I_{3} \star Q_{5} \star G_{4} ) \star
  ( H_{2} \star C_{1} \star W_{2} )\),

 \(A_{6} =
  ( X_{3} \star Y_{3} \star Z_{3} ) \star
  ( I_{3} \star Q_{5} \star G_{4} ) \star
  ( H_{3} \star C_{1} \star W_{2} )\),

 \(A_{7} =
  ( X_{3} \star Y_{3} \star Z_{3} ) \star
  ( I_{1} \star Q_{1} \star G_{4} ) \star
  ( H_{2} \star C_{1} \star W_{2} )\),

 \(A_{8} =
  ( X_{3} \star Y_{3} \star Z_{3} ) \star
  ( I_{1} \star Q_{1} \star G_{4} ) \star
  ( H_{3} \star C_{1} \star W_{2} )\).

 In Fig. 7 and Fig. 8,
 supersolution and subsolution are depicted.

\begin{center}
\begin{picture}(73,26)

\put(20,00){\makebox(0,0)[bl] {Fig. 7. Supersolution}}



\put(05,19){\circle*{1.5}} \put(13,19){\circle*{1.5}}
\put(21,19){\circle*{1.5}} \put(29,19){\circle*{1.5}}
\put(37,19){\circle*{1.5}} \put(45,19){\circle*{1.5}}
\put(53,19){\circle*{1.5}} \put(61,19){\circle*{1.5}}
\put(69,19){\circle*{1.5}}

\put(04,21){\makebox(0,0)[bl]{\(X\)}}
\put(12,21){\makebox(0,0)[bl]{\(Y\)}}
\put(20,21){\makebox(0,0)[bl]{\(Z\)}}
\put(28,21){\makebox(0,0)[bl]{\(I\)}}
\put(36,20.5){\makebox(0,0)[bl]{\(Q\)}}
\put(44,21){\makebox(0,0)[bl]{\(G\)}}
\put(52,21){\makebox(0,0)[bl]{\(H\)}}
\put(60,21){\makebox(0,0)[bl]{\(C\)}}
\put(68,21){\makebox(0,0)[bl]{\(W\)}}


\put(05,13){\oval(07,12)}

\put(13,13){\oval(07,12)}

\put(21,13){\oval(07,12)}

\put(29,13){\oval(07,12)}

\put(37,13){\oval(07,12)}

\put(45,13){\oval(07,12)}

\put(53,13){\oval(07,12)}

\put(61,13){\oval(07,12)}

\put(69,13){\oval(07,12)}

\put(03,14){\makebox(0,0)[bl]{\(X_{2}\)}}
\put(03,10){\makebox(0,0)[bl]{\(X_{3}\)}}

\put(11,14){\makebox(0,0)[bl]{\(Y_{2}\)}}
\put(11,10){\makebox(0,0)[bl]{\(Y_{3}\)}}

\put(19,14){\makebox(0,0)[bl]{\(Z_{2}\)}}
\put(19,10){\makebox(0,0)[bl]{\(Z_{3}\)}}

\put(27,14){\makebox(0,0)[bl]{\(I_{1}\)}}
\put(27,10){\makebox(0,0)[bl]{\(I_{3}\)}}

\put(35,14){\makebox(0,0)[bl]{\(Q_{1}\)}}
\put(35,10){\makebox(0,0)[bl]{\(Q_{5}\)}}

\put(43,14){\makebox(0,0)[bl]{\(G_{4}\)}}

\put(51,14){\makebox(0,0)[bl]{\(H_{2}\)}}
\put(51,10){\makebox(0,0)[bl]{\(H_{3}\)}}

\put(59,14){\makebox(0,0)[bl]{\(C_{1}\)}}

\put(66.5,14){\makebox(0,0)[bl]{\(W_{2}\)}}


\end{picture}
\end{center}

\begin{center}
\begin{picture}(73,26)

\put(21.5,00){\makebox(0,0)[bl] {Fig. 8. Subsolution}}



\put(05,19){\circle*{1.5}} \put(13,19){\circle*{1.5}}
\put(21,19){\circle*{1.5}} \put(29,19){\circle*{1.5}}
\put(37,19){\circle*{1.5}} \put(45,19){\circle*{1.5}}
\put(53,19){\circle*{1.5}} \put(61,19){\circle*{1.5}}
\put(69,19){\circle*{1.5}}

\put(04,21){\makebox(0,0)[bl]{\(X\)}}
\put(12,21){\makebox(0,0)[bl]{\(Y\)}}
\put(20,21){\makebox(0,0)[bl]{\(Z\)}}
\put(28,21){\makebox(0,0)[bl]{\(I\)}}
\put(36,20.5){\makebox(0,0)[bl]{\(Q\)}}
\put(44,21){\makebox(0,0)[bl]{\(G\)}}
\put(52,21){\makebox(0,0)[bl]{\(H\)}}
\put(60,21){\makebox(0,0)[bl]{\(C\)}}
\put(68,21){\makebox(0,0)[bl]{\(W\)}}


\put(05,13){\oval(07,12)}

\put(13,13){\oval(07,12)}

\put(21,13){\oval(07,12)}

\put(29,13){\oval(07,12)}

\put(37,13){\oval(07,12)}

\put(45,13){\oval(07,12)}

\put(53,13){\oval(07,12)}

\put(61,13){\oval(07,12)}

\put(69,13){\oval(07,12)}

\put(43,14){\makebox(0,0)[bl]{\(G_{4}\)}}

\put(59,14){\makebox(0,0)[bl]{\(C_{1}\)}}

\put(66.5,14){\makebox(0,0)[bl]{\(W_{2}\)}}


\end{picture}
\end{center}

 The obtained subsolution contains three elements
 (this combination will be considered as ``system kernel'').
 Thus, the aggregation process
 is considered as multiple choice problem
 for selection of DAs for subsystem
 \( \Theta = X \star Y \star Z \star I \star Q \star H \)
  (Fig. 9)
  (without taking into account compatibility).
  Corresponding binary variables are:
  \(\{x_{ij}\}, ~  i=\overline{1,6}, ~ j=\overline{1,2} \).

\begin{center}
\begin{picture}(74,29)

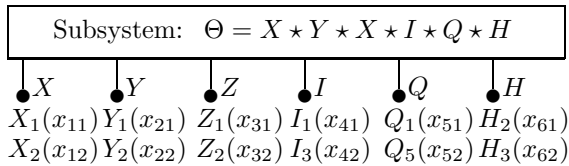
\put(6.5,00){\makebox(0,0)[bl]{Fig. 9. Selection of DAs for
subsystem}}

\put(00,09){\makebox(0,8)[bl]{\(X_{1}(x_{11})\)}}
\put(00,05){\makebox(0,8)[bl]{\(X_{2}(x_{12})\)}}

\put(12.5,09){\makebox(0,8)[bl]{\(Y_{1}(x_{21})\)}}
\put(12.5,05){\makebox(0,8)[bl]{\(Y_{2}(x_{22})\)}}

\put(25,09){\makebox(0,8)[bl]{\(Z_{1}(x_{31})\)}}
\put(25,05){\makebox(0,8)[bl]{\(Z_{2}(x_{32})\)}}

\put(37.5,09){\makebox(0,8)[bl]{\(I_{1}(x_{41})\)}}
\put(37.5,05){\makebox(0,8)[bl]{\(I_{3}(x_{42})\)}}

\put(50,09){\makebox(0,8)[bl]{\(Q_{1}(x_{51})\)}}
\put(50,05){\makebox(0,8)[bl]{\(Q_{5}(x_{52})\)}}

\put(62.5,09){\makebox(0,8)[bl]{\(H_{2}(x_{61})\)}}
\put(62.5,05){\makebox(0,8)[bl]{\(H_{3}(x_{62})\)}}

\put(02,15){\line(0,1){4}} \put(2,14){\circle*{1.7}}

\put(14.5,15){\line(0,1){4}} \put(14.5,14){\circle*{1.7}}

\put(27,15){\line(0,1){4}} \put(27,14){\circle*{1.7}}

\put(39.5,15){\line(0,1){4}} \put(39.5,14){\circle*{1.7}}

\put(52,15){\line(0,1){4}} \put(52,14){\circle*{1.7}}

\put(64.5,15){\line(0,1){4}} \put(64.5,14){\circle*{1.7}}

\put(02,14){\makebox(0,8)[bl]{~\(X\)}}
\put(14.5,14){\makebox(0,8)[bl]{~\(Y\)}}
\put(27,14){\makebox(0,8)[bl]{~\(Z\)}}
\put(39.5,14){\makebox(0,8)[bl]{~\(I\)}}
\put(52,13.5){\makebox(0,8)[bl]{~\(Q\)}}
\put(64.5,14){\makebox(0,8)[bl]{~\(H\)}}


\put(00,19){\line(1,0){74}} \put(00,26){\line(1,0){74}}
\put(00,19){\line(0,1){7}} \put(74,19){\line(0,1){7}}

\put(06,21){\makebox(0,8)[bl]{Subsystem:~
  \( \Theta = X \star Y \star X  \star I \star Q \star H  \) }}

\end{picture}
\end{center}

 Thus, the problem is:
 \[ \arg \min_{M \in \overline{D}}
 ~~
  \sum_{(i,j) \in S=\{(i,j)| x_{ij}=1\}} ~ | \delta (M, e_{ij}) |,\]
 \[ s.t. ~~ \sum_{i=1}^{6} \sum_{j=1}^{2}  a_{ij} x_{ij} \leq b;
 ~~ \sum_{j=1}^{2} x_{ij} =  1;
%
%
 ~~ x_{ij} \in \{0, 1\}.\]
 Estimates are presented in Table 6
 (illustrative, expert judgment).
 Some examples of the resultant solutions are:

 (1) \(b=42\):~
 \(x_{12}=1\)   (\(X_{3} \)),
 \(x_{22}=1\)   (\(Y_{3} \)),
 \(x_{32}=1\)   (\(Z_{3} \)),
 \(x_{41}=1\)   (\(I_{1} \)),
 \(x_{51}=1\)   (\(Q_{1} \)),
 \(x_{62}=1\)   (\(H_{3} \)),
 \(\Theta_{1}=X_{3}\star Y_{3}\star  Z_{3}\star I_{1}
 \star  Q_{1}\star H_{3} \),
   \(e(\Theta_{1})=(0,2,1,0)\);

 (2) \(b=53\):~
 \(x_{11}=1\)   (\(X_{2} \)),
 \(x_{21}=1\)   (\(Y_{2} \)),
 \(x_{32}=1\)   (\(Z_{3} \)),
 \(x_{41}=1\)   (\(I_{1} \)),
 \(x_{51}=1\)   (\(Q_{1} \)),
 \(x_{62}=1\)   (\(H_{3} \)),
 \(\Theta_{2}=X_{2}\star Y_{2}\star  Z_{3}\star I_{1}
 \star  Q_{1}\star H_{3} \),
   \(e(\Theta_{2})=(1,2,0,0)\);

 (3) \(b=87\):~
 \(x_{11}=1\)   (\(X_{2} \)),
 \(x_{21}=1\)   (\(Y_{2} \)),
 \(x_{31}=1\)   (\(Z_{2} \)),
 \(x_{42}=1\)   (\(I_{3} \)),
 \(x_{52}=1\)   (\(Q_{5} \)),
 \(x_{61}=1\)   (\(H_{2} \)),
 \(\Theta_{3}=X_{2}\star Y_{2}\star  Z_{2}\star I_{3}
 \star  Q_{5}\star H_{2} \),
   \(e(\Theta_{3})=(2,1,0,0)\).

\begin{center}
\begin{picture}(55,67)
\put(0.7,63){\makebox(0,0)[bl]{Table 6. Estimates for
aggregation}}

\put(00,0){\line(1,0){55}} \put(00,51){\line(1,0){55}}
\put(00,61){\line(1,0){55}}

\put(00,0){\line(0,1){61}} \put(20,00){\line(0,1){61}}
\put(43,0){\line(0,1){61}} \put(55,0){\line(0,1){61}}

\put(01,56.7){\makebox(0,0)[bl]{Selection of}}
\put(01,53.5){\makebox(0,0)[bl]{DA}}

\put(23,57.2){\makebox(0,0)[bl]{Multiset  }}
\put(23,52.5){\makebox(0,0)[bl]{estimate \(e_{ij}\)}}

\put(45.5,57){\makebox(0,0)[bl]{Cost}}
\put(45.5,52.2){\makebox(0,0)[bl]{(\(a_{ij}\))}}


\put(01,46){\makebox(0,0)[bl]{\(x_{11}\)}}
\put(07,45.4){\makebox(0,0)[bl]{\((X_{2})\)}}

\put(01,42){\makebox(0,0)[bl]{\(x_{12}\)}}
\put(07,41.4){\makebox(0,0)[bl]{\((X_{3})\)}}

\put(01,38){\makebox(0,0)[bl]{\(x_{21}\)}}
\put(07,37.4){\makebox(0,0)[bl]{\((Y_{2})\)}}

\put(01,34){\makebox(0,0)[bl]{\(x_{22}\)}}
\put(07,33.4){\makebox(0,0)[bl]{\((Y_{3})\)}}

\put(01,30){\makebox(0,0)[bl]{\(x_{31}\)}}
\put(07,29.4){\makebox(0,0)[bl]{\((Z_{2})\)}}

\put(01,26){\makebox(0,0)[bl]{\(x_{32}\)}}
\put(07,25.4){\makebox(0,0)[bl]{\((Z_{3})\)}}

\put(01,22){\makebox(0,0)[bl]{\(x_{41}\)}}
\put(07,21.4){\makebox(0,0)[bl]{\((I_{1})\)}}

\put(01,18){\makebox(0,0)[bl]{\(x_{42}\)}}
\put(07,17.4){\makebox(0,0)[bl]{\((I_{3})\)}}

\put(01,14){\makebox(0,0)[bl]{\(x_{51}\)}}
\put(07,13.4){\makebox(0,0)[bl]{\((Q_{1})\)}}

\put(01,10){\makebox(0,0)[bl]{\(x_{52}\)}}
\put(07,09.4){\makebox(0,0)[bl]{\((Q_{5})\)}}

\put(01,06){\makebox(0,0)[bl]{\(x_{61}\)}}
\put(07,05.4){\makebox(0,0)[bl]{\((H_{2})\)}}

\put(01,02){\makebox(0,0)[bl]{\(x_{62}\)}}
\put(07,01.4){\makebox(0,0)[bl]{\((H_{3})\)}}



\put(24,45.5){\makebox(0,0)[bl]{\((2,1,0,0)\)}}

\put(47,46){\makebox(0,0)[bl]{\(11\)}}


\put(24,41.5){\makebox(0,0)[bl]{\((0,2,1,0)\)}}

\put(48,42){\makebox(0,0)[bl]{\(4\)}}


\put(24,37.5){\makebox(0,0)[bl]{\((2,1,0,0)\)}}

\put(47,38){\makebox(0,0)[bl]{\(10\)}}


\put(24,33.5){\makebox(0,0)[bl]{\((0,1,1,1)\)}}

\put(48,34){\makebox(0,0)[bl]{\(2\)}}


\put(24,29.5){\makebox(0,0)[bl]{\((2,1,0,0)\)}}

\put(47,30){\makebox(0,0)[bl]{\(12\)}}


\put(24,25.5){\makebox(0,0)[bl]{\((0,2,1,0)\)}}

\put(48,26){\makebox(0,0)[bl]{\(6\)}}


\put(24,21.5){\makebox(0,0)[bl]{\((1,2,0,0)\)}}

\put(48,22){\makebox(0,0)[bl]{\(7\)}}


\put(24,17.5){\makebox(0,0)[bl]{\((3,0,0,0)\)}}

\put(47,18){\makebox(0,0)[bl]{\(20\)}}


\put(24,13.5){\makebox(0,0)[bl]{\((2,1,0,0)\)}}

\put(47,14){\makebox(0,0)[bl]{\(14\)}}


\put(24,09.5){\makebox(0,0)[bl]{\((3,0,0,0)\)}}

\put(47,10){\makebox(0,0)[bl]{\(21\)}}


\put(24,05.5){\makebox(0,0)[bl]{\((2,1,0,0)\)}}

\put(47,06){\makebox(0,0)[bl]{\(13\)}}


\put(24,01.5){\makebox(0,0)[bl]{\((0,2,1,0)\)}}

\put(48,02){\makebox(0,0)[bl]{\(5\)}}


\end{picture}
\end{center}


\section{Conclusion}

 The paper describes a hierarchical approach to
 composition of modular telemetry systems.
 Hierarchical morphological multicriteria design and
 multiple choice knapsack problem
 with interval multiset estimates are used for
 combinatorial synthesis and improvement
  of the telemetry system.
 Evidently,
 usage of more complicated assessment problems
 (e.g., \(P^{6,5}\)) will lead to more exact and realistic
 solving processes.

 In the future, it may be prospective
  to consider the following research directions:
 {\it 1.} examination of
  design and improvement/adaptation problems
 for telemetry systems
 as real-time reconfiguration;
 {\it 2.} examination of
 a distributed telemetry system
 that is based on a set of
 vehicles and/or a set of ground points;
 {\it 3.} examination  of various applications
 in engineering and management;
 {\it 4.} usage of interval multiset estimates
 for compatibility between design alternatives
 (in this case the combinatorial synthesis problem
 may be more easy);
 {\it 5.} consideration  of AI techniques
 (e.g., \cite{sabin98}); and
 {\it 6.} usage of the described approach
  in engineering/management/CS education.

%

\end{document}